\documentclass[preprint,aps,12pt]{revtex4}
\usepackage{graphicx,amsmath,amssymb}

\newcommand{\bea}{\begin{eqnarray}}
\newcommand{\eea}{\end{eqnarray}}

\newcommand{\be}{\begin{equation}}
\newcommand{\ee}{\end{equation}}

\renewcommand{\thesection}{\Roman{section}}
\renewcommand{\theequation}{\arabic{section}.\arabic{equation}\alph{eqnalpha}}

\newcounter{eqnalpha}

\newcommand{\balpha}[1]{\begin{eqnarray}\label{#1}\setcounter{eqnalpha}{1}}
\newcommand{\ealpha}{\end{eqnarray}\setcounter{eqnalpha}{0}}

\begin{document}

\begin{titlepage}

\title{Newtonian Kinetic Theory and the Ergodic-Nonergodic
Transition}

\author{Shankar P. Das$^{2,1}$ and Gene F. Mazenko$^{1}$}

\affiliation{$^1$The James Franck Institute and the Department of
Physics,
The University of Chicago, Chicago, Illinois 60637, U.S.A.\\
$^2$School of Physical Sciences, Jawaharlal Nehru University, New
Delhi - 110067, India.}

\vspace*{1cm}

\begin{abstract}
In a recent work we have discussed how kinetic theory, the
statistics of classical particles obeying Newtonian dynamics, can be
formulated as a field theory. The field theory can be organized to
produce a self-consistent perturbation theory expansion in an
effective interaction potential. In the present work we use this
development for investigating ergodic-nonergodic (ENE) transitions
in dense fluids. The theory is developed in terms of a core problem
spanned by the variables $\rho$, the number density, and $B$, a
response density. We set up the perturbation theory expansion for
studying the self-consistent model which gives rise to a ENE
transition. Our main result is that the low-frequency dynamics near
the ENE transition is the same for Smoluchowski and Newtonian
dynamics.  This is true despite the fact that term by term in a
density expansion the results for the two dynamics are fundamentally
different.
\end{abstract}


\maketitle
\end{titlepage}

\newpage

\section{Introduction}
\label{sec_intdn}

In previous work \cite{NDYN1}, referred to as ND1 from hereon, we
showed how the Newtonian dynamics (ND) of simple fluids can be
treated using field-theoretic methods\cite{FTSPD}. Here we apply
these methods to systems in equilibrium with an emphasis on the
ergodic-nonergodic (ENE) transition which is generally associated
with the mode coupling models\cite{bongsoo,wizland,szammel} for
glassy dynamics. We analyze the long-time behavior of the
density-density time correlation function. For high densities we
find that there is a slow dynamical regime which is very similar to
that found in the case of Smoulokowski dynamics
(SD)\cite{SDENE,MMS}. In the shorter-time regimes the SD dynamics is
considerably simpler than ND. The complications in the ND case arise
because of the unwieldy forms of the three-point vertices even for
the ideal-gas limit.  In the long-time limit we find that the
three-point cumulants and three-point vertices satisfy
fluctuation-dissipation relations which are very helpful in treating
the late-stage kinetics at high densities.

After determining the nature of the fluctuation dissipation
relations\cite{FDRSD} for the two-point cumulants and vertices, we derive an
exact expression (kinetic equation) satisfied by the density-density
correlation function. The form of this kinetic equation is the same
as found in projection operator treatments\cite{POT} of the problem.  This
approach involves the introduction of memory functions.  The
significant difference between our work here and the projection
operator method is that we obtain expressions for the memory
functions which are suitable for perturbation theory analysis. It is
not expressed in term of "projected dynamics". Instead the memory
function is expressed as the "BB" component of a two-point vertex
function  $\Gamma_{BB}$ associated with what we call a core problem.
The core problem involves the kinetics of the particle density
$\rho$ and a response field "B".

We then carry out perturbation theory in terms of a pseudo potential
and find at second order  that the memory function kernal $K_{BB}$
in the low-frequency regime is proportional to a quadratic-form in
the full density-density correlation function.  This suggests a
non-linear feedback loop similar to that found in mode-coupling
theory\cite{MCT1,MCT2} and in treating SD\cite{SDENE}.

When we focus on long-time solutions we find that the developments
for SD and ND are the same. In SDENE\cite{SDENE} we analyzed the leading
contribution near an ENE transition. In SM\cite{SM} it is shown the
ENE-transition is accompanied with a two-step decay process. SM shows
that there are a pair of exponents, $a$ and $b$, which characterize
the slow-time power-law solutions. They treat in detail the case of
hard spheres at second order in perturbation theory. We find here
that SD and ND systems share the same ENE transition according to
our results at second order in perturbation theory.  This point is
amplified in Ref.\onlinecite{GLD}  where higher order terms in perturbation
theory are treated. We speculate that there is a whole class of
systems which share the same statics and ENE transition.

\setcounter{equation}{0}
\section{Field-theoretical Formulation}
\label{sec_ftform}

\subsection{Newtonian Equations of Motion}
\label{sec_ftform1}

We discuss the kinetics satisfied by a system of $N$ particles with
mass $m$, position $R_{i}$ and momentum $P_{i}$ we consider
Newtonian dynamics, with equations of motion \be m\dot{R}_{i}=P_{i}
\ee \be \dot{P}_{i}=f_{i} ~~~, \ee where the force is given by \be
f_{i}=-\nabla_{i}U ~~~, \ee the total potential energy is \be
U=\frac{1}{2}\sum_{i,j}V(R_{i}-R_{j}) \ee where $V(R_{i}-R_{j})$ is
a standard pair potential between the $i$-th and $j$-th particles.
Starting with these equations of motion we can reexpress the problem
in terms of a path-integral formulation.

\subsection{Path-Integral Formulation}
\label{sec_ftform2}

In FTSPD and ND1 we introduced a field-theoretical formulation for
classical many-particles dynamics. Here we summarize the results for
Newtonian dynamics.  The grand partition function for a {\it core}
set of dynamical fields $\{\Phi_\alpha\}\equiv{\Phi}$ is given by

\be Z_{T}=\sum_{N=0}^{\infty}\frac{z^{N}}{N!}\mathrm{Tr} e^{-{\cal
A} +H\cdot\Phi} \label{eq:zT} \ee

\noindent where the trace is over the set of phase-space variables
$R_{i}(t)$ and $P_{i}(t)$, and the conjugate MSR\cite{MSR} variables
$\hat{R}_{i}(t)$ and $\hat{P}_{i}(t)$. $z$ is the fugacity and
$H_{i}$ is an external time-dependent field that couples to the set
of dynamical variables $\Phi_\alpha$. The MSR action ${\cal A}$ is
given by

\be \label{msraction}
 {\cal A} = \sum_i \int \Bigg [ i\hat{R}_i(t) \Big \{
\dot{R}_i(t)-\frac{P_i(t)}{m} \Big \} + i\hat{P}_i(t)\Big \{
\dot{P}_i(t)- f_i(t) \Big \} \Bigg ] dt ~~~. \ee

\noindent In the present work the set $\{\Phi_\alpha\}$ stands for
the core variables $\{\rho,B\}$ which are collective properties and
are expressed as a sum of one point functions.  Explicitly the
particle density is given by

\be \label{rho-defn}
 \rho (1)=\sum_{i = 1}^{N}\delta (x_{1} -
R_{i}(t_{1})), \ee

\noindent  and the response  field $B$:

\be \label{B-defn}
 B(x,t)=i\sum_{i=1}^N \hat{P}_i \cdot
 \frac{\partial}{\partial{{R}_i}}\delta(x-R_i(t))
~~~. \ee

\noindent The $B$-field is important in the present analysis and is
somewhat unfamiliar. We note that $B$ is the longitudinal
component of the vector field $\hat{\bf g} (x,t)$. The latter is
expressed as a collective density corresponding to the hatted MSR
field $\hat{\bf P}_i$ (with a factor $i$ ) conjugate to the momentum
${\bf P}_i$ of the $i$-th particle.

\begin{equation}
\label{Bfdt2} \hat{\bf g}(x,t)=\sum_{i=1}^N i\hat{\bf P}
\delta(x-R_i(t))
\end{equation}

\noindent  This $\hat{\bf g}(x,t)$ however is not the field
conjugate to momentum density ${\bf g}(x,t)$ used in the usual MSR
formulation of equations of nonlinear fluctuating hydrodynamics. The
hatted fields in the MSR field theory are generally viewed as
convenient  mathematical tools for ensuring the that the stochastic
equations of the dynamics are satisfied in the field theory. These
auxiliary fields give rise in a natural way to a set response
functions which are related to correlation functions through
suitable fluctuation dissipation relations. From such
considerations, the $B$ is best interpreted here as a suitable mean
to obtain linear fluctuation dissipation relations facilitating the
analysis of the dynamics developed here.

In general it is assumed that we can express $\Phi_\alpha$ as a sum
of one particle contributions

\be \label{defn-Phi}
\Phi_{\alpha}=\sum_{i=1}^{N}\phi_{\alpha}^{(i)}~~, \ee

\noindent where with the roman index $i$ we indicate that the
corresponding quantity is a single-particle  property. In the above
equation we have used a compact notation where the index $\alpha$
labels space, time and fields $\rho$ or $B$. We maintain the
notation from  here on. For $\alpha\equiv\rho$ we obtain the
collective density for which we have

\be \label{spd4}
\phi_{\rho}^{(i)}(1)=\delta (x_{1}-R_i(t_1))~~. \ee

\noindent For $\alpha\equiv{B}$ field, the corresponding
$\phi_{B}^{(i)}(1)$ is strongly dependent on the type of dynamics.
For the SD case\cite{FTSPD}

\be \phi_{B}^{(i)}(1)=D[i\hat{R}_i(t_1)\nabla_{x_1}+
\frac{1}{2}\nabla_{x_1}^2]\delta (x_1-R_i(t_1)) \ee

\noindent while  for the ND\cite{NDYN1} case the corresponding
result is

\be \phi_{B}^{(i)}(1)=-i\hat{P}_i(t_1)\nabla_{x_1} \delta
(x_1-R_i(t_1)) \ee

\noindent where $\hat{P}_i(t_1)$ is the MSR conjugate momentum for
the selected particle.

For our purposes the MSR action ${\cal A}$ defined in eqn.
(\ref{msraction}) can be written as a sum of two parts,

\be \label{eq:mac}
 {\cal A} = {\cal A}_0 +{\cal A}_I ~~.\ee

\noindent  The first term on the RHS ${\cal A}_0$ is the
noninteracting MSR action. The interacting part of the action is
given by

\be {\cal A}_I=\frac{1}{2}\sum_{\alpha,\nu}\int d1d2 \Phi
_{\alpha}(1)\sigma _{\alpha\nu} (12)\Phi _{\nu}(2) \label{eq:26} \ee

\noindent where the Greek labels range over  $\rho$ and $B$. The
interaction matrix, for systems in equilibrium in the distant past,
is defined just as in SD, by

\be \sigma_{\alpha \beta} (12)=V(12) \left[
\delta_{\alpha{B}}\delta_{\beta\rho }
+\delta_{\alpha\rho}\delta_{\beta{B}}\right] \label{eq:27} \ee

\noindent and

\be V(12)=V(x_{1}-x_{2})\delta (t_{1}-t_{2}) ~~~. \ee

\noindent Notice that the response field $B$ is chosen such that the
interaction part of the action has the form given by
Eq.(\ref{eq:26}).

\noindent For studying the dynamics of a many-particle system, we
will be interested in the cumulants generated by corresponding
generating functional

\be W[H]=\ln Z_{T}[H]~~. \ee

\noindent The one-point average of a field $\Phi_\alpha$ defined in
eqn. (\ref{defn-Phi}) in an external field is obtained in terms of
the functional derivative

\be \label{eq:Hf}  G_{\alpha}=\langle \Phi_{\alpha}\rangle
=\frac{\delta}{\delta H_{\alpha}}W[H]~~. \ee

\noindent The full cumulants $G_{\alpha\beta}$,
$G_{\alpha\beta\gamma}$ etc. are defined by successive functional
derivatives of $G_\alpha$,

\bea \label{twptcum}
 G_{\alpha\beta} &=& \frac{\delta}{\delta H_{\beta}}G_{\alpha} \\
 \label{thptcum}
G_{\alpha\beta\gamma}&=& \frac{\delta}{\delta H_{\gamma}}G_{\alpha\beta}  \\
G_{\alpha\beta\ldots \delta} &=& \frac{\delta}{\delta H_{\alpha}}
\frac{\delta}{\delta H_{\beta}}.....\frac{\delta}{\delta H_{\delta}}
W[H] ~~.\eea

\noindent We also need to deal with the {\it single-particle}\cite{SCF}
quantities

\be {\cal G}_{\alpha\beta\ldots \delta}=\Big \langle\sum_{i
=1}^{N}\phi^{(i)}_{\alpha}
\phi^{(i)}_{\beta}\ldots\phi^{(i)}_{\delta}\Big \rangle~~~.
\label{eq:2.23} \ee

\setcounter{equation}{0}
\section{Fluctuation Dissipation Relations}
\label{sec_fdt}

In this section we explore the nature of fluctuation-dissipation
relations in case of Newtonian dynamics assumed to be in
equilibrium. The correlation  functions between the fields in the
filed theory is primarily determined with the corresponding action
functional of the associated field theory. The MSR action without
initial conditions and external fields is given by Eq.
(\ref{msraction}). We now explore the invariance of the action
${\cal A}$ under certain symmetry operations. In a stationary state
we have time-translational invariance and time-reversal symmetry.

\subsection{Symmetry transformations}
\label{sec_fdt2}

\noindent 1.  The time-reversal symmetry $T$ is defined as,

\bea
T R_i(t) &=& R_i(-t) \nonumber \\
T\hat{R}_i(t) &=& -\hat{R}_i(-t)\nonumber \\
T P_i(t) &=& -P_i(-t) \nonumber \\
T\hat{P}_i(t) &=& \hat{P}_i(-t) ~~.
\label{ABL-trans}
\eea

\noindent We consider how the MSR-action ${\cal A}$ changes under
time-reversal

\bea {\cal A}^{'} &=& T  {\cal A} = \sum_i \int dt \Bigg[ i\left \{
-\hat{R}_i(-t) \right \} \left \{-
\frac{{\partial}R_i(-t)}{\partial(-t)}
+\frac{P_i(-t)}{m} \right \} \nonumber \\
&+& i \left \{\hat{P}_i(-t) \nonumber \right \}
\left \{ \frac{{\partial}P_i(-t)}{\partial(-t)}-f_i(-t) \right \}
\Bigg ] \nonumber \\
\eea

\noindent Letting $t\rightarrow -t$ in the time integral, we obtain
${\cal A}'=T{\cal A}={\cal A}$ if the limits of integration are
symmetric. This includes $t_{2}\rightarrow\infty$ and
$t_{1}\rightarrow -\infty$ here. We conclude that the MSR action
remains invariant under time-reversal.

\noindent 2. In ND1 we have discussed the Fluctuation-Dissipation
symmetry (FDS) for the field theory corresponding to the MSR action
functional defined in Eq. (\ref{msraction}). It was demonstrated \cite{ABL}
that the following transformation $\tau$ keeps the MSR action
invariant :

\bea \label{FD-tr1}
\tau R_i(t) &=& R_i(-t) \\
\label{FD-tr2}
\tau\hat{R}_i(t) &=& -\hat{R}_i(-t) - i\beta f_{i}(-t) \\
\label{FD-tr3}
\tau P_i(t) &=& -P_i(-t) \\
\label{FD-tr4} \tau\hat{P}_i(t) &=& \hat{P}_i(-t) - i
\frac{\beta}{m}P_i(-t)~~. \label{ABL-trans} \eea

\noindent We have therefore $\tau {\cal A}= {\cal A}$.

\noindent 3. The inversion symmetry denoted by $\tau_1$ is defined
as

\bea
\tau_{1} R_i(t) &=& -R_i(t) \nonumber \\
\tau_{1}\hat{R}_i(t) &=& -\hat{R}_i(t) )\nonumber \\
\tau_{1} P_i(t) &=& -P_i(t) \nonumber \\
\tau_{1}\hat{P}_i(t) &=& -\hat{P}_i(t) ~~.
\label{ABL-trans}
\eea

\noindent The following invariant properties of the Action ${\cal
A}$ easily follows : $\tau_{1}{\cal A} = {\cal A}$.

\noindent 4. Finally we define  the symmetry $\tau_2$ as

\bea
\tau_{2} R_i(t) &=& R_i(t) \nonumber \\
\tau_{2}\hat{R}_i(t) &=& -\hat{R}_i(t)
\nonumber \\
\tau_{2} P_i(t) &=& P_i(t) \nonumber \\
\tau_{2}\hat{P}_i(t) &=& -\hat{P}_i(t) ~~.
\label{ABL-trans}
\eea

\noindent The following invariant properties of the Action
$\tau_{2}{\cal A}^{*} = {\cal A}$ is satisfied.

\subsection{Fluctuation-Dissipation symmetry}
\label{sec_fdt3}

From the definitions (\ref{rho-defn}) and (\ref{B-defn}) it follows
that under the transformation $\tau$ the core fields $\rho (x,t)$
and $B(x,t)$  change as

\bea \label{Brho1} \tau \rho(x,t) &=& \rho(x,-t) \\
\label{Bfdt3} \tau B(x,t) &=& B(x,-t)-\beta
\frac{\partial\rho(x,-t)}{\partial t} \eea

\noindent In ND1 we had demonstrated the following Fluctuation
dissipation theorem involving two-point correlation functions
between the $\rho$ and $B$ fields.

\begin{equation}
\label{Bfdt4} G_{fB}(t-t')=\theta
(t-t'){\beta}\frac{\partial}{\partial{t}}{G_{f\rho}}(t-t'),
\end{equation}

\noindent for any function $f[\rho]$.

Here we are interested in higher-order FDR which are most easily
expressed in terms of Fourier transforms. Consider the mixed
correlation function

\be \label{mixcor}
 C_{BB...B\rho...\rho}(12...\ell \ell +1...n)=\Big
\langle B(1)B(2)...B(\ell) \rho(\ell +1)...\rho (n) \Big \rangle
~~~.
\ee

\noindent where

\be \rho (j)=\rho (q_{j},\omega_{j})=\int dt e^{i\omega_{j}t}
\sum_{i=1}^{N}e^{-iq_{j}\cdot R_{i}(t)} \ee

\noindent and

\be B(j)=B(q_{j},\omega_{j})=\int dt e^{i\omega_{j}t}(-q_j)\cdot
\sum_{i=1}^{N}\hat{P}_{i}(t)e^{-iq_{j}\cdot R_{i}(t)} \ee

\noindent Under the transformation $\tau$ we have

\bea \tau B(1) &=& B(q_{1},-\omega_{1}) +i\beta \omega_{1}\rho
(q_{1},-\omega_{1}) \\
\tau \rho (1) &=& \rho (q_{1},-\omega_{1})~~. \eea

\noindent Hence the mixed correlation defined above in Eq.
(\ref{mixcor}) is obtained as

\bea
 C_{BB...B\rho...\rho}(12...\ell \ell +1...n) &=& \Big \langle \tau
B(1)\tau B(2)...\tau B(\ell) \tau \rho(\ell +1)...\tau \rho (n)\Big
\rangle \nonumber \\
&=& \Big \langle [ B(\tilde{1})+i\beta \omega_{1}\rho (\tilde{1})]
 [B(\tilde{2})+i\beta \omega_{2}\rho (\tilde{2})] \nonumber \\
&..& [B(\tilde{\ell})+i\beta \omega_{\ell}\rho (\tilde{\ell})]
 \rho (\tilde{\ell +1})... \rho (\tilde{n})\Big \rangle
\eea

\noindent where $\tilde{j}=q_{j},-\omega_{j}$. Multiplying these
out, each correlation funcion has arguements with tildes.  Using the
result \be \tau_{1}\tau_{2}{\cal A}^{*}={\cal A} \ee and \be
\tau_{1}\tau_{2}B^{*}(1)=B(\tilde{1}) \ee \be
\tau_{1}\tau_{2}\rho^{*}(1)=\rho(\tilde{1}) \ee we find the simple
result

\bea C_{BB...B\rho...\rho}(12...\ell \ell +1...n) &=& \Big [ \Big
\langle [B(1)-i\beta \omega_{1}\rho (1)]
 [B(2)-i\beta \omega_{2}\rho (2)] \nonumber \\
 &..& [B(\ell)-i\beta \omega_{\ell}\rho (\ell)]...
 \rho (\ell +1)...
 \rho (n) \Big \rangle {\Big ]}^*~~~.
\label{eq:3.24} \eea

\noindent If we define

\be \psi(i)=B(i)-i\beta\omega_{i}\rho (i)~~, \ee

\noindent the mixed correlation function is obtained as

\be C_{BB...B\rho\rho}(12\ldots\ell,\ell+1\ldots n) = \Big \langle
{\psi}(1) {\psi}(2) \ldots {\psi}(\ell) {\rho}(\ell+1) \ldots
{\rho}(n) \Big \rangle \ee

\noindent  One easily sees that this holds for the cumulants\cite{FDRSD}

\be G_{BB...B\rho\rho}(12\ldots\ell,\ell+1\ldots n)
=G^{*}_{\psi\psi...\psi\rho\ldots\rho}(12\ldots\ell,\ell+1\ldots n)
\ee

\noindent and for the single-particle quantities

\be {\cal G}_{BB...B\rho\rho}(12\ldots\ell,\ell+1\ldots n)
=\tilde{\cal
G}^{*}_{\psi\psi...\psi\rho\ldots\rho}(12\ldots\ell,\ell+1\ldots n)
~~~.
\ee

\subsubsection{Two-point cumulants}

Using the FD relations at the two-point level we find

\be
G_{B\rho}(12)=G^{*}_{B\rho}(12)
+i\beta\omega_{1}G^{*}_{\rho\rho}(12) ~~~. \label{gbr-0} \ee

\noindent Due to translational invariance in space and time

\bea
\label{gbr-1}
G_{\alpha\beta}(12) &=& G_{\alpha\beta}(1)\delta(1+2) \\
\label{gbr-2}
G_{\rho{B}}(1) &=& G^{*}_{B\rho}(1)~~~. \eea

\noindent  We see that Eq.(\ref{gbr-0}) can be written as

\be G_{B\rho}(1)=G_{\rho{B}}(1) +i\beta\omega_{1}G^{*}_{\rho\rho}(1)
~~~. \label{gbr-3} \ee

\noindent Since $G_{\rho\rho}(1)$ is real, using eqns. (\ref{gbr-2})
and (\ref{gbr-3}), we have the conventional result

\be Im G_{B\rho}(1)= -Im G_{\rho
B}(1)=\frac{\beta\omega_{1}}{2}G_{\rho\rho}(1)
\ee

\noindent which can be used to construct the dispersion relation

\be \label{gbr-e1}
 G_{\rho B}(1)=\int\frac{d\bar{\omega}}{2\pi}
\frac{\beta\bar{\omega}G_{\rho\rho}(q_{1},\bar{\omega})}
{\omega_{1}-\bar{\omega}+i\eta}~~~. \ee

\noindent Taking the inverse time Fourier transform we find

\be \label{gbr-te1} G_{\rho B}(q,t)=\theta
(t)\beta\frac{\partial}{\partial t} G_{\rho\rho}(q,t)~~~. \ee

\noindent Using the FD relation (\ref{gbr-e1}) we obtain for
$G_{\rho{B}}(q,0)$

\be \label{gbr-e11}
 G_{\rho B}(q,\omega=0)=\int\frac{d\bar{\omega}}{2\pi}
\frac{\beta\bar{\omega}G_{\rho\rho}(q_{1},\bar{\omega})}
{-\bar{\omega}+i\eta} = -\beta \bar{\rho} S(q) \ee

\noindent where $S(q)$ is the static structure factor at wave vector
$q$ and $\bar{\rho}$ is the average density so that the RHS
represents the correlation of density fluctuations at equal time.
From eqn. (\ref{gbr-3}), we obtain the very useful result

\be \label{2p-disp} \-i\beta \omega G_{\rho\rho}(q,\omega )+
G_{\rho B}(q,\omega )-G_{B\rho }(q,\omega ) =0 \ee

\noindent Eqn. (\ref{2p-disp}) is written in a compact form as
${\cal F}[G]=0$, where the functional ${\cal F}$ acting on
the elements of a $2\times2$ matrix $A_{\mu\nu}$ is defined as follows

\be \label{fdr-func} {\cal F} \left [ A \right ] \equiv \mathrm{Tr}
\left [ (i\beta \omega {\cal I} +\varepsilon ) \cdot A(q,\omega )
\right]~~. \ee

\noindent We have introduced the traceless anti-symmetric matrix
$\varepsilon$ in the RHS of eqn. ({fdr-func}) :

\be \label{eps-matrix} \varepsilon_{B\rho}=1,\ \  \varepsilon_{\rho
B}=-1,\ \  \varepsilon_{\mu\mu}=0. \ee

\noindent  ${\cal I}$ is the $2\times{2}$ identity matrix. The
functional $\tilde{\cal F}$ is defined as

\be \label{fdr-func2} \tilde{\cal F} \left [ A \right ] \equiv
\mathrm{Tr} \left [ (-i\beta \omega {\cal I} +\tilde{\varepsilon} )
\cdot A(q,\omega )\right] \ee

\noindent where the transpose matrix $\tilde{\varepsilon}_{\mu\nu}=
\varepsilon_{\nu\mu}$. Vanishing of the above defined functional
${\cal F}[A]$ then ensures the fluctuation-dissipation relation
(FDR) among the of the elements of the matrix $A$. It is
straightforward to show that the relation

\be \label{cfdrmp}
 {\cal F}[A]=0 \ee

\noindent  is satisfied for the two-point propagator $A$ belonging
to any member of the set $\{G^{(0)},G,{\cal G} \}$. With the
condition (\ref{cfdrmp}) the elements of $A$  will be referred to as
FDR matrix propagator (FDRMP). We show in Appendix \ref{appC} how to
build new propagators which also obey the same FDR.  If
$A_{\alpha\beta}(q,\omega )$  and $D_{\alpha\beta}(q,\omega )$ are
respectively two FDRMP then the matrix $H_{\alpha\beta}$ defined in
terms of the  interaction matrix $\sigma_{\mu\nu}$   as

\be H_{\alpha\beta}(q,\omega)=
\sum_{\delta\gamma}A_{\alpha\delta}(q,\omega)\sigma_{\delta\gamma}(q)
D_{\gamma\beta}(q,\omega) \ee

\noindent is also a FDRMP since it satisfies the relation ${\cal
F}[H_{\alpha\beta}]=0$ . This property of composite propagators
following the same Fluctuation-dissipation relation is an important
ingredient in the analysis of memory functions discussed later in
this work.

\subsubsection{Two-point Vertices}

The two-point vertices $\Gamma_{\alpha\beta}(1)$ are defined as the
corresponding elements of the inverse of the general two-point
correlation matrix $G_{\alpha\beta}$ :

\be \label{tver-def}
\sum_{\mu}\Gamma_{\alpha\mu}(1)G_{\mu
\beta}(1)=\delta_{\alpha\beta} ~~~. \ee

\noindent For $\alpha=\rho$ and $\beta=B$, using $G_{BB}(1) = 0$\cite{GBB}, it
follows from the above equation that $\Gamma_{\rho\rho}(1)=0$. For
other choices of $\alpha,\beta$ we obtain,

\bea
\label{gamtw3}\Gamma_{\rho B}(1)G_{B\rho}(1) &=& 1 \\
\label{gamtw4}\Gamma_{B\rho }(1)G_{\rho B}(1) &=& 1 \\
\label{gamtw5}\Gamma_{B B}(1)G_{B\rho}(1) +\Gamma_{B
\rho}(1)G_{\rho\rho}(1) &=& 0 ~~~. \label{eq:3.43}
\eea

\noindent Using Eqs. (\ref{gamtw3}) and (\ref{gamtw4}),  Eq.
(\ref{gamtw5}) can be written in the form

\be \label{gamtw}G_{\rho\rho}(1)=-G_{\rho B}(1) \Gamma_{B
B}(1)G_{B\rho}(1) ~~~. \ee

\noindent We easily find

\be \label{gamtw6} \Gamma_{B\rho}(q,\omega ) =\Gamma^{*}_{\rho
B}(q,\omega ) \ee

\noindent Starting with Eq.(\ref{gbr-0}):

\be \label{gamtw7} G_{B\rho}(1)=G_{B\rho}^{*}(1) +i\beta
\omega_{1}G_{\rho\rho}^{*}(1) \ee

\noindent we can use Eq.(\ref{gamtw3}), (\ref{gamtw4})  and
(\ref{gamtw6}) to write

\be \frac{1}{\Gamma_{\rho B}(1)}= \frac{1}{\Gamma_{B\rho }(1)}
+i\beta \omega_{1}G_{\rho\rho}(1) \label{eq:118} \ee

\noindent Furthermore using Eq.(\ref{gamtw5}) and
Eq.(\ref{eq:118}), and canceling a common denominator, gives

\be \label{twovf-fdt}
 i\beta\omega_{1}\Gamma_{BB}(1)
=\Gamma_{\rho B}(1) -\Gamma_{B\rho}(1) ~~~. \ee

\noindent In terms of the functional ${\cal F}$ defined in Eq.
(\ref{fdr-func}) the above relation is obtained as

\be \label{FDR-Gamma} {\cal F}[\Gamma]=0~~, \ee

\noindent where the matrix vertex $\Gamma_{\alpha\beta}$ is
defined in the $\{\rho,B\}$ domain.

\subsubsection{FD symmetry and Three-point cumulants}

There are substantial differences in specific cumulants for the
different realizations of simple-fluid dynamics. For example the
noninteracting three-point cumulant $G^{(0)}_{B\rho\rho}(123)$ for
SD and ND are quite different. Despite this fact we find that both sets
of cumulants satisfy the same fluctuation dissipation relations.

In treating the three-point cumulants
involving the fields at three different points, we adopt the
following notation (in Fourier space)

\bea \label{eq:1}
G_{\rho_i\rho_j\rho_k}&\equiv&{G_{\rho\rho\rho}(ijk)}=
{G_{\rho\rho\rho}(jki)}={G_{\rho\rho\rho}(kji)} \\
G_{\rho_i\rho_jB_k}&\equiv&{G_{\rho\rho{B}}(ijk)}
={G_{\rho{B}\rho}(jki)}={G_{{B}\rho\rho}(kij)},\\
G_{B_iB_jB_k}&\equiv&{G_{BBB}(ijk)}=G_{BBB}(jki) \ \  \mathrm{etc.}
\eea

\noindent where the set $\{ijk\}\Longrightarrow\{123\},\{231\}$, and
$\{321\}$. The factor $\beta$ will for now be absorbed in the
frequency $\omega$.  For cumulants involving three
$B$ fields we obtain using the basic
fluctuation-dissipation relation (\ref{gbr-0}) the following
relations :

\bea \label{eBBB} G_{B_iB_jB_k} &=& G^*_{B_iB_jB_k} +
i\omega_iG^*_{\rho_i
B_jB_k} +i\omega_jG^*_{B_i\rho_j B_k} +i\omega_k G^*_{B_iB_j\rho_k}\\
&-& {\omega_i\omega_j}{G^*_{\rho_i\rho_j B_k}} -
{\omega_j\omega_k}{G^*_{B_i \rho_j \rho_k}} -
{\omega_k\omega_i}{G^*_{\rho_k B_j B_i}} - i\omega_i\omega_j\omega_k
{G^*_{\rho_i\rho_j \rho_k}} \nonumber \eea

\noindent  Since $G_{BBB}=G_{BBB}^{*}=0$, using this in the complex
conjugate of Eqn. (\ref{eBBB}) gives the relation

\bea \label{G3e0}
 0 &=& i\omega_iG_{\rho_i B_jB_k} +i\omega_jG_{B_i\rho_j
B_k} +i\omega_k G_{B_iB_j\rho_k} +{\omega_i\omega_j}{G_{\rho_i\rho_j
B_k}} \\
&+& {\omega_j\omega_k}{G_{B_i \rho_j \rho_k}} +
{\omega_k\omega_i}{G_{\rho_k B_j \rho_i}} -i\omega_i\omega_j\omega_k
{G_{\rho_i\rho_j \rho_k}} \nonumber \eea

\noindent Similarly for the cumulants respectively involving two and
one $B$ fields are obtained as follows :

\bea \label{eBBr} G_{B_iB_j\rho_k} &=& G^*_{B_iB_j\rho_k} +
i\omega_iG^*_{\rho_i B_j\rho_k} +i\omega_jG^*_{B_i\rho_j \rho_k}
-{\omega_i\omega_j}{G^*_{\rho_i\rho_j \rho_k}} \\
\label{eBrr} G_{B_i\rho_j\rho_k} &=& G^*_{B_i\rho_j\rho_k} +
i\omega_iG^*_{\rho_i \rho_j\rho_k} ~~~.\eea

\noindent To simplify the above relations involving the three-point
cumulants, we define $\mu_i=i\omega_iG_{\rho_iB_jB_k}$,
$\sigma_i=\omega_j\omega_k{G_{B_i\rho_j\rho_k}}$, and
$\nu=\omega_i\omega_j\omega_k{G_{\rho_i\rho_j\rho_k}}$. In terms of
the newly defined quantities we rewrite the Eqs.
(\ref{eBBr})-(\ref{G3e0}) respectively as

\bea -\mu^*_k &=& \mu_k+\sigma_j+\sigma_i- i\nu \\
\sigma_i &=& \sigma^*_i+i\nu \\
\mu_i&+&\mu_j+\mu_k+\sigma_i+\sigma_j+\sigma_k = i\nu \eea

\noindent where $\nu$ is real. Solving the above equations we obtain

\bea
\sigma^{''}_i &=& \nu/2 \\
\mu^{'}_i &=&-\frac{\sigma^{'}_j+\sigma^{'}_k}{2}\\
\mu^{''}_T &=& \frac{\nu}{2} \eea

\noindent where $\mu_T=\mu_i+\mu_j+\mu_k$ is the sum of the $\mu$'s.
These results seem rather general.

\subsubsection{FD symmetry and Three-point Cumulants}

Next we analyze the relations between the different three-point
vertex functions. For this we use the general definition of the
three-point cumulants and the corresponding three-point
vertex function as,

\be \label{G3v3} G_{\alpha_i{\mu_j}\nu_k} =
-G_{\alpha\alpha^\prime}(i)G_{\mu\mu^\prime}(j)G_{\nu\nu^\prime}(k)
\Gamma_{{\alpha_i^\prime}{\mu_j^\prime}{\nu_k^\prime}}\ee

\noindent where $\alpha_i$ stands for the field $\alpha$ at the
point $i$ and so on. The fields $\alpha$,$\mu$ and $\nu$ are
respectively taken from the set $\{\rho,B\}$. In the RHS of Eq.
(\ref{G3v3}) we have used the notation $G_{\mu\nu}(i)$ following
Eq. (\ref{gbr-0}), {\em i.e.},
 $G_{\mu\nu}(ij)=G_{\mu\nu}(i)\delta(i+j)$. Since $G_{BB}=0$,
 taking all the three fields to be $B$ at the respective points
 $i$, $j$, and $k$ we obtain, ,

\be \label{G3e4} G_{B_iB_jB_k}=
-G_{B\rho}(i)G_{B\rho}(j)G_{B\rho}(k)\Gamma_{\rho_i\rho_j\rho_k} \ee

\noindent Since $G_{B_iB_jB_k}=0$ it follows from the last equation
that $\Gamma_{\rho_i\rho_j\rho_k}=0$. The expression (\ref{G3v3})
for the various three-point cumulants in terms of two-point
cumulants simplifies due to the vanishing of
$\Gamma_{\rho\rho\rho}$. Let us first consider the expression
(\ref{G3e0}) involving {\em three} $B$ fields. In the appendix
\ref{appA} we show that by a) using the expression (\ref{G3v3})  for
$G_{\alpha\mu\nu}$ in Eq. (\ref{G3e0}), and b) by collecting the
various coefficients of the different vertices $\Gamma$'s, we obtain
the following relation between the three-point vertices.

\bea \label{vcond1} && i\Big ( \omega_i\Gamma_{B_i\rho_j\rho_k}+
\omega_j \Gamma_{\rho_iB_j\rho_k} + \omega_k
\Gamma_{\rho_i\rho_jB_k} \Big ) -
i \omega_i\omega_j\omega_k\Gamma_{B_iB_jB_k} \nonumber \\
&+& \Big \{ {\omega_i\omega_j}\Gamma_{B_iB_j\rho_k}+
{\omega_j\omega_k}\Gamma_{\rho_iB_jB_k} +
{\omega_k\omega_i}\Gamma_{B_i\rho_jB_k} \Big \}=0 ~~~.\eea

\noindent Next, we consider the relations (\ref{eBBr}) which contain
three-point cumulants involving correlation of {\em two} $B$ fields.

 \be \label{eBBr1} G^*_{B_iB_j\rho_k} = G_{B_iB_j\rho_k} -
i\omega_iG_{\rho_i B_j\rho_k} -i\omega_jG_{B_i\rho_j \rho_k}
-{\omega_i\omega_j}{G_{\rho_i\rho_j \rho_k}}\ee

\noindent Using the the three-point cumulants corresponding to this
case in terms of vertex functions as defined in Eq. (\ref{G3v3})
and collecting the coefficients of the various vertex functions, we
obtain in Appendix \ref{appA} the result :

\be \label{eBBr9} {i\omega_k}\Gamma^*_{\rho_i \rho_j B_k }+
i\omega_i \Gamma_{B_i\rho_j\rho_k} +i\omega_j
\Gamma_{\rho_iB_j\rho_k} + \omega_i\omega_j \Gamma_{B_iB_j\rho_k} =0
~~~.
\ee

\noindent By doing circular permutations of the indices $\{i,j,k\}$
in Eq. (\ref{eBBr9}) we obtain the following relations involving
the vertices with one $B$-field leg:

\bea \label{eBBr10} {i\omega_i}\Gamma^*_{B_i\rho_j \rho_k}+
i\omega_j \Gamma_{\rho_iB_j\rho_k} +i\omega_k
\Gamma_{\rho_i\rho_jB_k} + \omega_j\omega_k \Gamma_{\rho_iB_jB_k}
&=& 0 \\
\label{eBBr11} {i\omega_j}\Gamma^*_{\rho_i B_j \rho_k}+ i\omega_k
\Gamma_{\rho_i\rho_jB_k} +i\omega_i \Gamma_{B_i\rho_j\rho_k} +
\omega_k\omega_i \Gamma_{B_i\rho_jB_k} &=& 0 \eea

\noindent The above relations are further simplified with the
following notations :

\bea \label{sreln1}
\phi_i &=& i\omega_{i}\Gamma_{B_i\rho_j\rho_k},\\
\label{sreln2}
\psi_{i} &=& \omega_{j}\omega_{k}\Gamma_{\rho_i B_jB_k},\\
\label{sreln3} P&=&
-i\omega_{i}\omega_{j}\omega_{k}\Gamma_{B_iB_jB_k}~~. \eea

\noindent  We obtain from Eq. (\ref{vcond1}) using the above
notation the result

\be P+\phi_{T}+\psi_{T}=0 \label{viden1} \ee

\noindent where $\phi_{T}=\phi_{1}+\phi_{2}+\phi_{3}$ and
$\psi_{T}=\psi_{1}+\psi_{2}+\psi_{3}$. In the simplified notation
the Eqs. (\ref{eBBr9}) - (\ref{eBBr11}) respectively reduces to the
cyclic forms

\be \label{viden2}
 -\phi_{i}^{*}+\phi_{j}+\phi_{k}+\psi_{i}= 0~~. \ee

\noindent Looking at the imaginary parts of these equations we have
$\psi_{i}''=-\phi_{T}''$, while for the real parts $
-\phi_{i}'+\phi_{j}'+\phi_{k}'+\psi_{i}' = 0$. The last equation has
the simple solution

\be \phi_{j}' = -\frac{1}{2}(\psi_{i}'+\psi_{k}') \ee

\noindent In terms of the original notation, using the defining
relation (\ref{G3v3}) for the different vertices, we obtain the
following set of {\em nonperturbative} relations among the various
vertex functions.

\bea \Gamma_{B_i\rho_j\rho_k}'' &=& \frac{1}{2}
(\omega_{k}\Gamma_{B_i\rho_j B_k}'+\omega_{j} \Gamma_{B_iB_j\rho_k
}') \nonumber \\
\omega_{j}\omega_{k}\Gamma_{\rho_iB_jB_k}'' &=&
\omega_{i}\omega_{k}\Gamma_{B_i\rho_jB_k}''
=\omega_{i}\omega_{j}\Gamma_{B_iB_j\rho_k}'' \nonumber \\
=-\Big ( \omega_i\Gamma_{B_i\rho_j\rho_k}' &+&
\omega_j\Gamma_{\rho_iB_j\rho_k}' +\omega_{k}\Gamma_{\rho_i\rho_j
B_k}' \Big )~~. \eea

\noindent The vertex function $\Gamma_{B_iB_jB_k}$ is real, {\em
i.e.}, $\Gamma_{B_iB_jB_k}'' = 0$. The real part satisfies the
relation

\bea \omega_{i}\omega_{j}\omega_{k}\Gamma_{B_iB_jB_k}' &=& -2\Big [
\omega_i\Gamma_{B_i\rho_j\rho_k}' +\omega_j\Gamma_{\rho_iB_j\rho_k}'
+\omega_k\Gamma_{\rho_i\rho_jB_k}' \Big ] ~~~.\eea

\setcounter{equation}{0}
\section{Kinetic Equations}
\label{sec_kineq}

We now present starting from the basic equation (\ref{tver-def}),
the derivation of a kinetic equation for time evolution of the
density autocorrelation function  $G_{\rho\rho}(q,t)$. This kinetic
equation involves the different contributions to the vertex
functions $\Gamma_{ij}$. The analysis is similar to the discussions
in SDENE\cite{SDENE} and MMS\cite{MMS}. These developments start
with a discussion of the form of $\Gamma_{\alpha\beta}(q,\omega )$
in the time domain. If we look at

\be \label{twovert} \Gamma_{B\rho}(q, \omega)=\frac{1}{G_{\rho
B}(q,\omega)}~~. \ee

\noindent The response function $G_{\rho{B}}$ vanish algebraically
as $\omega\rightarrow\infty$, and hence the two-point vertex
$\Gamma_{B\rho}$ diverges in this limit.  We assume that

\be \lim_{\omega\rightarrow\infty} \Gamma_{B\rho}(q,
\omega)=-D_{q}\omega^{2}+i\omega A_{q}+C_{q} \ee

\noindent where the coefficients $A$, $C$ and $D$ are real and not
universal. We define the "local" quantity

\be \Gamma^{(\ell )}_{B\rho}(q, \omega)=-D_{q}\omega^{2}+i\omega
A_{q}+C_{q} \label{gam_l} \ee

\noindent for all frequencies. The corresponding subtracted
quantities are obtained as

\be \label{gam_t}
\Gamma^{(s)}_{B\rho}(q, \omega)= \Gamma_{B\rho}(q,
\omega)- \Gamma^{(\ell )}_{B\rho}(q, \omega) \ee

\noindent which vanish for large frequencies.  At low frequencies

\be \Gamma^{(s )}_{B\rho}(q,\omega=0)= \Gamma_{B\rho}(q, \omega=0)-
\Gamma^{(\ell )}_{B\rho}(q, \omega=0 )~~. \ee

\noindent Using the Eqs. (\ref{gbr-e11}) and  (\ref{twovert}), we
obtain $\Gamma_{B\rho}(q, 0)=-{[\bar{\rho} \beta S(q)]}^{-1}$. From Eq.
(\ref{gam_l}) it follows that $\Gamma^{(\ell )}_{B\rho}(q, 0
)=C_{q}$ and hence using Eq. (\ref{gam_t}) we obtain

\be \label{gam_s} \Gamma^{(s )}_{B\rho}(q, \omega=0)=
-C_q-{[\bar{\rho} \beta S(q)]}^{-1}~~. \ee

\noindent Next,  assuming  that the FDR holds locally, we obtain
from Eq. (\ref{gam_l}) and FD relation (\ref{twovf-fdt})

\be \omega \Gamma^{(\ell )}_{BB}(q,\omega ) =
-2\omega A_{q} \ee

\noindent Hence $\Gamma^{(\ell )}_{BB}(q, 0 )= -2 \beta^{-1} A_{q}$.
Assuming $A_{q}\geq 0$ and taking inverse Fourier transforms we
obtain in the time space the following results.

\bea \label{bb-lt}
\hat{\Gamma}^{(\ell )}_{BB}(q,t-t' )
&=& -\beta^{-1} A_{q}\delta (t - t') \\
\label{brho-lt} \hat{\Gamma}^{(\ell )}_{B\rho}(q,t-t' ) &=&
\left[D_{q}\frac{\partial^{2}}{\partial t ^{2}}
-A_{q}\frac{\partial}{\partial t} +C_{q}\right]\delta (t-t') \eea

\noindent where the $\Gamma$'s in the time space are denoted with a
hat. For the full two-point vertex functions $\Gamma_{B\rho}$ and
$\Gamma_{BB}$ the following relations involving the local and
subtracted parts,

\bea \hat{\Gamma}_{B\rho}(q,t-t' ) &=&
\hat{\Gamma}^{(\ell)}_{B\rho}(q,t-t' )+
\hat{\Gamma}^{(s)}_{B\rho}(q,t-t' ) \\
\hat{\Gamma}_{BB}(q,t-t' ) &=& -\beta^{-1}A_{q}\delta (t-t' )+
\hat{\Gamma}^{(s )}_{BB}(q,t-t' ) ~~~. \eea

\noindent It then follows for the subtracted parts
$\Gamma^{(s)}_{B\rho}$ that we have the dispersion relation ( since
the FDR holds locally for $\Gamma^{(\ell)}_{B\rho}$ )

\be \label{gamfd1}
\Gamma^{(s)}_{B\rho }(q,\omega )=
\int\frac{d\bar{\omega}}{2\pi}
\frac{\beta\bar{\omega}\Gamma^{(s)}_{BB}(q,\bar{\omega} )} {\omega
-\bar{\omega}+i\eta} \ee

\noindent reflecting the fact that the vertex function
$\Gamma_{B\rho}$ is analytic in the upper half plane. From the above
FDR it follows directly that

\be \label{gamfd2} \Gamma^{(s)}_{B\rho }(q,\omega=0 )=
\beta\hat{\Gamma}^{(s)}_{BB}(q,0) ~~~. \ee

\noindent In the time domain the FDR (\ref{gamfd1}) reads

\be \label{gamfd3} \hat{\Gamma}^{(s)}_{B\rho }(q,t-t')=\beta\theta
(t-t') \frac{\partial}{\partial t} \hat{\Gamma}^{(s)}_{BB}(q,t-t' )
~~~. \ee

\subsection{Memory Function Equation}
\label{sec_kineq1}

The density auto correlation function $G_{\rho\rho}$ is of
particular interest in the present theoretical model for studying
the slow dynamics of a dense liquid. The natural order parameter for
the ENE transition which is the primary focus of this paper is the
long-time limit of this function. In this section we obtain an
equation of motion for the density correlation function. First we
express Eq.(\ref{eq:3.43}) in the time-domain

\be \label{memeq1} \hat{\Gamma}^{(\ell
)}_{B\rho}(q,0)G_{\rho\rho}(q,t-t') +\hat{\Gamma}^{(\ell
)}_{BB}(q,0)G_{B\rho}(q,t-t') =\Psi_{q}(t,t') \ee where \be \Psi
(t,t')= -\int_{-\infty}^{t}ds
\hat{\Gamma}^{(s)}_{B\rho}(t-s)G_{\rho\rho}(s-t')
-\int_{-\infty}^{t'}ds \hat{\Gamma}^{(s)}_{BB}(t-s)G_{\rho
B}(t'-s)~~. \ee

\noindent In writing Eq. (\ref{memeq1})  we have used the fact that
the response functions are time ordered so that
$\hat{\Gamma}^{(s)}_{B\rho}(t-s)\sim \theta (t-s)$ and $G_{\rho
B}(t'-s)\sim\theta (t'-s)$.  We then use the fluctuation dissipation
relations (\ref{gbr-te1}) and (\ref{gamfd3}) to obtain

\be -\Psi (t,t') = \int_{-\infty}^{t}ds
\left[\frac{\partial}{\partial s} \beta
\hat{\Gamma}^{(s)}_{BB}(t-s)\right] G_{\rho\rho}(s-t')+
\int_{-\infty}^{t'}ds \hat{\Gamma}^{(s)}_{BB}(t-s)
\frac{\partial}{\partial s} \beta G_{\rho\rho}(t'-s) ~~~. \ee

\noindent Integrating by parts\cite{13} in the first integral gives

\bea -\Psi (t,t') &=&
\beta\hat{\Gamma}^{(s)}_{BB}(0)G_{\rho\rho}(t-t')
-\int_{-\infty}^{t}ds\beta\hat{\Gamma}^{(s)}_{BB}(t-s)
\frac{\partial}{\partial
s}G_{\rho\rho}(s-t') \nonumber \\
&+&\int_{-\infty}^{t'}ds \beta\hat{\Gamma}^{(s)}_{BB}(t-s)
\frac{\partial}{\partial s} G_{\rho\rho}(t'-s) \nonumber \\
&=& \beta\hat{\Gamma}^{(s)}_{BB}(0)G_{\rho\rho}(t-t')
-\int_{t'}^{t}ds \beta\hat{\Gamma}^{(s)}_{BB}(t-s)
\frac{\partial}{\partial s} G_{\rho\rho}(t'-s) \eea

\noindent  where we have assumed that  $t > t'$.  We then have the
kinetic equation

\bea  \label{eq:101} \left [ \hat{\Gamma}_{B\rho}^{(\ell)}(q,0)
+\beta\hat{\Gamma}^{(s)}_{BB}(q,0) \right ] G_{\rho\rho}(q,t-t')
-\int_{t'}^{t}ds \beta\hat{\Gamma}^{(s)}_{BB}(q,t-s)
\frac{\partial}{\partial s} G_{\rho\rho}(q,t'-s)=0
~~~.\nonumber \\
\eea

\noindent Since $t > t'$ we have dropped the
$\hat{\Gamma}_{BB}^{(\ell)}(q,0)G_{B\rho}(q, t-t')$ term in the LHS
of Eq. (\ref{memeq1}). Using the relations (\ref{gamfd2}) and
(\ref{gam_s}) we obtain for the quantity within square brackets on
the LHS of (\ref{eq:101}) as

\bea \hat{\Gamma}_{B\rho}^{(\ell
)}(q,0)+\beta\hat{\Gamma}^{(s)}_{BB}(q,0)
&=&\hat{\Gamma}_{B\rho}^{(\ell )}(q,0)+\Gamma_{B\rho
}^{(s)}(q,\omega=0) ~~~. \eea

\noindent Using the relations (\ref{gam_s}) and (\ref{brho-lt})
respectively for the second and the first terms on the RHS, the
kinetic equation then takes the form

\be \label{keqn1} \left[D_{q}\frac{\partial^{2}}{\partial
t^{2}}-A_{q} \frac{\partial}{\partial
t}-\frac{1}{\bar{\rho}{\beta}S(q)}\right] G_{\rho\rho}(q,t)
+\int_{t'}^{t}ds \beta\hat{\Gamma}^{(s)}_{BB}(q,t-s)
\frac{\partial}{\partial s} G_{\rho\rho}(q,t'-s) =0 ~~~. \ee

\noindent We see that our dynamical problem is now in the form of a
memory function equation and the dynamic part of the memory function
is given by the subtracted part $\hat{\Gamma}^{(s)}_{BB}$ of the
vertex function. The static structure factor is the same for all
fluids with the same pair potential.  The coefficients $A_{q}$ and
$D_{q}$ can be constructed using perturbation theory. In the case of
ND  we have the ideal gas result:

\be \Gamma_{B\rho}^{(0)}(q,\omega )= -\frac{1 } {\beta\bar{\rho}{\cal
S}^{*}(z)}
\ee

where $z={\omega}/(\sqrt{2}qv_0)$,
$v_0=\sqrt{k_BT/m}$ is the thermal velocity and

\be {\cal S}(x)=1-2xe^{-x^2}\int_0^{x}du e^{u^2}-i\sqrt{\pi}
xe^{-x^2}~~. \ee

\noindent In the large frequency limit

\be \Gamma_{B\rho}^{(0)}(q,\omega )= -\frac{1 }
{\beta\bar{\rho}}\left[-\left(\frac{\omega}{q{v_0}}\right)^{2}+1\right]
\ee and on comparing this relation with Eq. (\ref{gam_l}) we can
identify at the lowest order,

\bea \label{keqn-c1} A_0(q) &=& 0, \\
\label{keqn-c2} C_0(q) &=& -\frac{1}{\beta\bar{\rho}}, \\
\label{keqn-c3} q^2D_0(q) &=& {\beta}mC_0(q) =-\frac{m}{\bar{\rho}} \eea

\noindent where the equilibrium density is given by $\bar{\rho}$. With
the above identification we obtain the kinetic Eq. (\ref{keqn1}) in the
form

\be \label{keqn2} \left[\frac{\partial^{2}}{\partial t^{2}}
+\Omega_q^2\right] G_{\rho\rho}(q,t) +\int_{0}^{t}ds
\hat{\Gamma}_{\mathrm{mc}}(q,t-s) \frac{\partial}{\partial s}
G_{\rho\rho}(q,t'-s) =0 ~~~. \ee

\noindent where we have defined

\bea \Omega_q^2 &=& \frac{q^2}{\beta{m}S(q)}\equiv{q^2}c^2(q) \\
\Gamma_{\mathrm{mc}} (q,t) &=&
-\frac{\beta\bar{\rho}}{m}q^2\hat{\Gamma}^{(s)}_{BB}(q,t). \eea

\noindent Note that this resembles the second-order oscillator
equation of standard MCT \cite{MCT1,MCT2} {\em without} the bare
dissipation coefficient. In the case of Newtonian dynamics the
microscopic dynamics is reversible and the kinetic equation
(\ref{keqn2}) has been obtained without projection in to a space of
slow modes. Indeed as we discuss below in order for the ENE
transition to take place we need to break the time-reversal
symmetry.

The kinetic equation is diagonalized using a Laplace transform
defined as

\be
\tilde{G}_{\rho\rho}(q,z)=-i\int_{0}^{\infty}dte^{izt}G_{\rho\rho}(q,
t) ~~~. \ee

\noindent  The Laplace transformed  kinetic equation (\ref{keqn1})
reduces to the form for the normalized density correlation function
$F(q,t)$

\be \label{nep-eq1} F(q,z)~=~
\frac{\tilde{G}_{\rho\rho}(q,z)}{\bar{\rho} S(q)}~=~
\frac{z+iq^2\Gamma_{\mathrm{mc}}(q,z)}
{z[z+iq^2\Gamma_{\mathrm{mc}}(q,z)]-\Omega_q^2} \ee

\noindent where $\bar{\rho} S(q)$ in the denominator of the LHS above is
the equal-time correlation function and normalize $F(q,t)$ to unity.
We obtain the following integral equation for $F(q,z)$

\bea \label{nep-integ} \frac{F(q,z)}{1-zF(q,z)}&=&
-\frac{iq^2}{\Omega_q^2(q)}\int_0^{\infty} dt e^{izt}
{\beta}\Gamma_{\mathrm{mc}}(q,t) \nonumber \\
&=& i\bar{\rho} S(q) \int_0^{\infty} dt e^{izt}
{\beta^2}\Gamma^{(s)}_{BB}(q,t) \nonumber \\
&=&  -\bar{\rho} S(q){\beta^2}\Gamma^{(s)}_{BB}(q,z)
~~~.\eea

\noindent We now turn to the mechanism which produces large $F(q,z)$
and ${\Gamma}^{(s)}_{BB}(q,z)$ as $z\rightarrow 0$. This involves
determining the memory function as a functional of $F(q,t)$. We
define $F(q)$ as the long time limit of the normalized density
correlation function $F(q,t)$ to denote the so called non ergodicity
parameter.

\be {\lim}_{z\rightarrow{0}} \{ zF(q,z) \} \rightarrow F(q) \ee

\noindent From Eq. (\ref{nep-integ}) we obtain the following
integral equation for $F(q)$

\bea \label{nep-eq2} \frac{F(q)}{1-F(q)} =  \bar{\rho} S(q)
{\beta^2}\Gamma^{(s)}_{BB}(q)~~.
\label{eq:4.35}
\eea

\noindent $\Gamma^{(s)}_{BB}(q)$ is a functional of $F(q)$

\be {\lim}_{z\rightarrow{0}} \{- z\Gamma^{(s)}_{BB}(q,z) \}
\rightarrow \Gamma^{(s)}_{BB}(q) \ee

\noindent and is obtained from the explicit evaluation of the
subtracted part of the vertex function $\Gamma_{BB}$ in a
perturbation series. Solution of the resulting self-consistent
equation, Eq.(\ref{nep-eq2}), for $F(q)$ corresponds to what we call
the ENE problem. This now primarily involves expressing
$\Gamma^{(s)}_{BB}(q,t)$ self consistently in terms of the
correlation functions. The self-consistent treatment of the dynamics
constitutes the basic feedback mechanism and gives rise to the slow
dynamics characteristics of the mode coupling theories\cite{MCT1,MCT2}.

\setcounter{equation}{0}
\section{Self-Consistent Development}
\label{sec_scdev}

\subsection{General}
\label{sec_scdev1}

In Ref. 1 we obtained that the canonical partition function for the
$N$ particle classical system following Newtonian dynamics (ND) in
the form

\be Z_{N}[H,h,\hat{h}] =\int \prod_{i=1}^{N}{\cal D}(\Psi_{i}){\cal
D}(\hat{\Psi}_{i}) {\cal D}(\Psi_{i}^{(0)}) e^{-{\cal A}_0-{\cal
A}_I+H\cdot\phi+h\cdot \psi +\hat{h}\cdot\hat{\psi}} \label{eq:24}
\ee

\noindent where $\{\Psi_i,\hat{\Psi}_i\}$ are respectively the
microscopic variables for the $i$-th particle ($i=1,...,N$) and its
hatted counterpart in a Martin Siggia Rose (MSR) field theoretic
formulation of the problem.  In the above expression for $Z_N$, the
superscript $0$ in $\Psi_i^{(0)}$ in the RHS refers to the initial
state of the system. The MSR action is written as a sum of two parts
as stated in eqn. (\ref{eq:mac}). ${\cal A}_0$ is the quadratic part
of the action including the quadratic contribution to the initial
probability distribution. ${\cal A}_I$ is the interaction part of
the MSR action given in eqn. (\ref{eq:26}). The set
$\{h_i,\hat{h}_i\}$ denote the currents conjugate to
$\{\Psi_i,\hat{\Psi}_i\}$ for $i=1,...,N$ respectively. The
collective field $\Phi_\alpha$  is defined in eqn. (\ref{defn-Phi})
and $H_\alpha$ is current conjugate to $\Phi_\alpha$. The canonical
partition function (\ref{eq:24}) can be written in the convenient
form

\be
 Z_{N}={\mathrm{Tr}}^{(N)} e^{-A_{I}+H\cdot\Phi}
\label{eq:23} \ee

\noindent where we have introduced the $N$ particle average as,

\be {\mathrm{Tr}}^{(N)} =\int \prod_{i=1}^{N}{\cal D}(\Psi_{i}){\cal
D}(\hat{\Psi}_{i}) d\Psi_{i}^{(0)} e^{-A_{0}} ~~~. \ee

\noindent The partition for the interacting system is obtained using
the identity

\be \label{spd1} e^{-A_I+H\cdot\Phi} = e^{\hat{A}_T}e^{H.\Phi} \ee

\noindent in terms of operator

\be \label{spd2} \hat{A}_T = \frac{1}{2} \int d1 \int d2
\sum_{\alpha,\beta} \hat{H}_\alpha(1)
\sigma_{\alpha\beta}(12)\hat{H}_\beta(2) \ee

\noindent where  $\hat{H}_\alpha(1)=
[\delta/\delta\hat{H}_\alpha(1)]$. We rewrite the partition function
in a form that allows us to formally carry out the average in eqn.
(\ref{eq:23})  using the functional identity (\ref{spd1}).

\be \label{spd3}
 Z_N = e^{\hat{A}_T} {\mathrm{Tr}}^{(N)} e^{H.\Phi}~~. \ee

\noindent The sum over the degrees of freedom in eqn. (\ref{spd4})
factorizes into a product of sums over the degrees of freedom of
each particle. Together, these observations lead to the following
result for the noninteracting Partition function,

\be \label{spd5}
 Z^{(0)}_N = {\mathrm{Tr}}^{(N)} e^{H.\Phi}={(Z_1)}^N~~. \ee

\noindent where $Z_1$ is the single particle partition function.

\be \label{spd6} Z_1 = {\mathrm{Tr}}^{(1)} e^{H.\Phi} \ee

\noindent Working in the grand canonical ensemble, the grand
partition function for the interacting problem given by
(\ref{eq:zT}) is obtained in the form

\be \label{eq:zT1} Z_{T}[H,h,\hat{h}] =
\sum_{N=0}^{\infty}\frac{\rho_0^N}{N!} Z_{N}[H,h,\hat{h}] =
e^{\hat{A}_T} e^{W_0}~~, \ee

\noindent in terms of the single particle function $W_0$. In eqn.
(\ref{eq:zT1}), $z=\rho_0$ is the fugacity. The cumulants of the
fields $\Phi_{\alpha}$  are generated by taking functional
derivatives of the generating functional $W[H]=\ln~ Z_{T}$ with
respect the fields $H$ introduced in eqn. (\ref{eq:Hf}) above.

It was shown in FTSPD that the one-point average $G_\alpha$ in a
field $\Phi_\alpha$  satisfies the fundamental identity

\be G_{\alpha}=\mathrm{Tr} \phi_{\alpha} e^{H\cdot\phi +\Delta W
[H]}. \label{eq:194} \ee

\noindent where $\mathrm{Tr}$ is the sum over all the degrees of
freedom including the conjugate MSR degrees of freedom for a {\em
single} particle. The functional $\Delta{W}$ is defined by

\be \Delta W[H] =W[H+F]-W[H] \ee

\noindent where

\be F_{\alpha}=\sum_{\beta}\sigma_{\alpha\beta}\phi_{\beta}~~.
\label{eq:45} \ee

\noindent The dependence of the theory  on the interaction potential
is controlled by $\Delta W[H]$. This quantity  is
expressed in a functional Taylor-series expansion in powers of the
potential $V$ as,

\be \Delta W[H] =\sum_{\alpha}F_{\alpha}\frac{\delta}{\delta
H_{\alpha}}W[H]
+\sum_{\alpha\beta}\frac{1}{2}F_{\alpha}F_{\beta}\frac{\delta^{2}}{\delta
H_{\alpha}\delta H_{\beta}}W[H]+\cdots \label{eq:45a} \ee

\noindent with $F_{\alpha}$ given by Eq.(\ref{eq:45}). For systems
fluctuating in equilibrium the force matrix is given in terms of
fourier transforms by eqn. (\ref{eq:27}) above,

\be \sigma_{\alpha\beta}(q)=V(q)[\delta_{\alpha \rho}
\delta_{\beta{B}}+\delta_{\alpha{B}}\delta_{\beta\rho}] ~~.\ee

\noindent The result {(\ref{eq:194})} was established in FTSPD (Ref
2.) using functional methods and ND1 (Ref. 1) using algebraic
methods.

We conveniently introduce the set of cumulants:

\be G_{\alpha\beta\ldots \delta}=\frac{\delta}{\delta
H_{\alpha}}\frac{\delta}{\delta H_{\beta}} \ldots
\frac{\delta}{\delta H_{\delta}} W[H] \label{fder-e1} \ee

\noindent to obtain the functional Taylor series expansion
(\ref{eq:45a}) as,

\be \Delta W[H] =\sum_{\alpha}F_{\alpha}G_{\alpha}
+\sum_{\alpha\beta}\frac{1}{2}F_{\alpha}F_{\beta}G_{\alpha\beta}
+\sum_{\alpha\beta\gamma}\frac{1}{3!}F_{\alpha}F_{\beta}F_{\gamma}
G_{\alpha\beta\gamma} +\ldots \label{fder-e2}\ee

\noindent  Clearly, in this form $\Delta W$ is treated as  a
functional  correlation functions $G_{\alpha}$, $G_{\alpha\beta}$,
$G_{\alpha\beta\gamma}$, ... and irreducible vertex functions.

We established in FTSPD a dynamic generalization of the static
Ornstein-Zernike relation\cite{OZR} given by eqn.(59) there.
Inserting Eq.(\ref{fder-e2}) in eqn. (\ref{fder-e1}) we obtain

\be G_{\alpha\beta}=\frac{\delta}{\delta H_{\beta}}G_{\alpha} ={\cal
G}_{\alpha\beta}+\sum_{\gamma}c_{\alpha\gamma}G_{\gamma\beta}
\label{eq:47} \ee

\noindent where ${\cal G}_{\alpha\beta}$ is given by
Eq.(\ref{eq:2.23}) and is related to the self1s- correlation
function:

\be {\cal G}_{\alpha\beta}=\mathrm{Tr}
\phi_{\alpha}\phi_{\beta}e^{H\cdot\phi +\Delta W}~~. \label{eq:54}
\ee

\noindent The memory function\cite{TDDC}, or dynamic direct
correlation function in eqn. (\ref{eq:47}) is obtained as

\be c_{\alpha\beta}=\mathrm{Tr} \phi_{\alpha}e^{H\cdot\phi +\Delta
W} \frac{\delta}{\delta G_{\beta}}\Delta W. \label{eq:34} \ee

\noindent Since $\Delta W$ can be treated as a functional of $G_{i}$
we see at this stage that we have available a self-consistent
theory. If we define the matrix-inverses

\bea \sum_{\delta}\Gamma_{\alpha\delta} G_{\delta\beta}
&=& \delta_{\alpha\beta} \label{eq:56} \\
\sum_{\delta}\gamma_{\alpha\delta}{\cal G}_{\delta\beta} &=&
\delta_{\alpha\beta} \label{eq:39} \eea

\noindent then the two-point vertex is given without approximation
by \be \Gamma_{\alpha\beta}=\gamma_{\alpha\beta}+K_{\alpha\beta}
\label{eq:50} \ee where \be
K_{\alpha\beta}=-\sum_{\delta}\gamma_{\alpha\delta}c_{\delta\beta}
\label{eq:38} \ee is the collective contribution to the vertex
function $\Gamma_{\alpha\beta}$. From eqn. (\ref{eq:34}) it follows
that we can obtain the $c_{\alpha\beta}$ and hence $K_{\alpha\beta}$
as an expansion in the interaction potential by expressing
$\Delta{W}$ in a similar series

\be \label{delWseries} \Delta{W}=
\Delta{W}^{(1)}+\Delta{W}^{(2)}+\Delta{W}^{(3)}+\cdots ~~\ee

\subsection{Collective part of vertex function at First-order}
\label{sec_scdev2}

 Going to first order in pertubation theory we
have \be \Delta W^{(1)}=\sum_{\alpha}F_{\alpha}G_{\alpha} ~~~. \ee
Then we need to compute \be f_{\alpha}^{(1)}= \frac{\delta}{\delta
G_{\alpha}}\Delta W^{(1)} =F_{\alpha} \ee which goes back into
Eq.(\ref{eq:34}) giving the result \be c_{\alpha\beta}=\mathrm{Tr}
\phi_{\alpha}F_{\beta}e^{\Delta W} =\sum_{\delta}{\cal
G}_{\alpha\delta}\sigma_{\delta\beta} ~~~. \label{eq:192} \ee
Putting Eq.(\ref{eq:192}) into Eq.(\ref{eq:38}) and using
Eq.(\ref{eq:39}) gives the very simple result

\be K^{(1)}_{\alpha\beta}=-\sigma_{\alpha\beta} ~~~. \ee

\noindent This result satisfies the FDR discussed below in a trivial
way: \be \label{mfnkb1} K_{\rho B}^{(1)}=K_{B\rho }^{(1)}=-\beta
V(q) \ee while \be K_{B B}^{(1)}=-\frac{2}{\beta \omega}Im ~K_{B\rho
}^{(1)}=0 ~~~. \ee

\subsection{Collective part of the vertex function
 at Second Order }
\label{sec_scdev3}

The  collective part $K$ of the two-point vertex function $\Gamma$,
at second order is denoted as $K^{(2)}$ and is obtained by
determining $f_{\alpha}^{(2)}$, defined as \be
f_{\alpha}^{(2)}=\frac{\delta}{\delta G_{\alpha}}\Delta W^{(2)} \ee
at second order.  We have \be \Delta
W^{(2)}=\frac{1}{2}\sum_{\alpha\beta}
F_{\alpha}F_{\beta}G_{\alpha\beta} \ee and \be
f_{\alpha}^{(2)}=\frac{\delta \Delta W^{(2)}}{\delta G_{\alpha}}
=\frac{1}{2}\sum_{\beta\delta}F_{\beta}F_{\delta}\frac{\delta
G_{\beta\delta}}{\delta G_{\alpha}} \label{eq:203} ~~~. \ee We then
have the standard functional manipulations,

\be \frac{\delta}{\delta G_{\delta}}G_{\alpha\beta}
=-\sum_{\nu\eta}G_{\alpha\nu}\frac{\delta{G^{-1}_{\nu\eta}}}{\delta
G_{\delta}} G_{\eta\beta}
=-\sum_{\nu\eta}G_{\alpha\nu}G_{\beta\eta}\Gamma_{\nu\eta\gamma} \ee

\noindent where in writing the last equation we introduce the
three-point vertex

\be \Gamma_{\alpha\beta\gamma}=\frac{\delta}{\delta
G_{\gamma}}\Gamma_{\alpha\beta} ~~~. \ee

\noindent Putting this back into Eq.(\ref{eq:203}) gives \be
f_{\alpha}^{(2)}=\frac{\delta \Delta W^{(2)}}{\delta G_{\alpha}}
=-\frac{1}{2}\sum_{\nu\mu\kappa\delta} F_{\nu}F_{\mu}
G_{\nu\kappa}G_{\mu\delta}\Gamma_{\kappa\delta\alpha}~~. \ee

\noindent This in turn goes into Eq.(\ref{eq:34}) and

\be c_{\alpha\beta}^{(2)}=-\frac{1}{2}\sum_{\kappa\gamma\delta\eta}
\mathrm{Tr} \phi_{\alpha} e^{\Delta W}F_{\kappa}F_{\gamma}
G_{\kappa\delta}G_{\gamma\eta}\Gamma_{\delta\eta\beta} \nonumber \ee
\be =-\frac{1}{2}\sum_{\delta\nu\kappa\eta\pi\varrho\theta} {\cal
G}_{\alpha\delta\nu} \sigma_{\delta\eta} \sigma_{\nu\pi}
G_{\eta\varrho}G_{\pi\theta}\Gamma_{\varrho\theta\beta} \nonumber
\ee

\noindent where

\be {\cal G}_{\alpha\beta\delta}=\mathrm{Tr}~
\phi_{\alpha}\phi_{\beta} \phi_{\delta}e^{\Delta W} \ee

\noindent is a three-point self-correlation involving the products
of three single particle quantities $\phi_\alpha^{(i)}$,
$\phi_{\beta}^{(i)}$ and  $\phi_\delta^{(i)}$. The second order
contributions to the two-point vertex function
$\Gamma_{\alpha\beta}$ can be written more symmetrically in terms of
the three-point self-vertex $\gamma_{\alpha\beta\delta}$. The latter
is defined as

\be {\cal G}_{\alpha\beta\gamma}=
-\sum_{{\alpha}'{\beta}'{\delta}'}{\cal G}_{\alpha\alpha^\prime}
{\cal G}_{\beta\beta^\prime}{\cal
G}_{\delta\delta^\prime}\gamma_{{\alpha^\prime}{\beta^\prime}{\delta^\prime}}
~~~. \label{eq:4.96} \ee

\noindent Then the collective part of the two-point-vertex is given
at second-order by

\bea K_{\alpha\beta}^{(2)} &=& -\frac{1}{2}
\sum_{\nu{\nu}'\mu{\mu}'\delta{\delta}'\eta{\eta}'\pi{\pi}'}
\gamma_{\alpha\nu} {\cal G}_{\nu{\nu}'}{\cal G}_{\mu{\mu}'} {\cal
G}_{\delta{\delta}'}\gamma_{{\nu}'{\mu}'{\delta}'} \sigma_{\mu
{\eta}'}\sigma_{\delta{\pi}'} G_{{\eta}'\eta}G_{{\pi}'\pi}
\Gamma_{\beta\eta\pi} \nonumber \\
&=& -\frac{1}{2}
\sum_{\nu{\nu}'\mu{\mu}'\delta{\delta}'\eta{\eta}'\pi{\pi}'}
\gamma_{\alpha{\nu}'{\delta}'} {\cal G}_{\nu{\nu}'} {\cal
G}_{\delta{\delta}'} \sigma_{\nu\eta}\sigma_{\delta\pi}
G_{\eta{\eta}'}G_{\pi{\pi}'}\Gamma_{{\eta}'{\pi}'\beta} \nonumber \\
&=&-\frac{1}{2} \sum_{\nu\mu\delta\eta\pi} \gamma_{\alpha\nu\delta}
\bar{G}_{\nu\mu}\bar{G}_{\delta\eta}\Gamma_{\mu\eta\beta}
\label{selfcol} \eea

\noindent where in writing the last equality we have introduced the
self-dressed propagator

\be \label{bgdef} \bar{G}_{\alpha\beta}= \sum_{\nu\delta} {\cal
G}_{\alpha\nu}\sigma_{\nu\delta}G_{\delta\beta}~~. \ee

\noindent In Appendix \ref{appC} we demonstrate that
$\bar{G}_{\alpha\beta}$ is a FDRMP with the same generic properties
as $G_{\alpha\beta}$.

From its definition the collective part of the two-point vertex can
be constructed to be symmetric and the second-order contribution can
be written in the symmetric form:

\bea
 K_{\alpha\beta}^{(2)} &=& -\frac{1}{2}
\sum_{\nu\delta\eta\theta}
\left[\gamma_{\alpha\nu\delta}\bar{G}_{\nu\eta}\bar{G}_{\delta
\theta}\Gamma_{\eta\theta\beta} +\gamma_{\beta
\nu\delta}\bar{G}_{\nu\eta}\bar{G}_{\delta\theta}\Gamma_{\eta\theta\alpha}\right]
\nonumber
\\&=& -\frac{1}{2}
\sum_{\nu\delta\eta\theta}
\left[\gamma_{\alpha\nu\delta}\bar{G}_{\nu\eta}\bar{G}_{\delta
\theta}\Gamma_{\eta\theta\beta}
+\Gamma_{\alpha\nu\delta}\bar{G}_{\nu\eta}\bar{G}_{\delta\theta}
\gamma_{\eta\theta\beta}\right] \nonumber \\
&=&  -\frac{1}{2} \sum_{\nu\delta\eta\theta} \Big
[(\Gamma_{\alpha\nu\delta}+\gamma_{\alpha\nu\delta})
\bar{G}_{\nu\eta}\bar{G}_{\delta\theta}
(\Gamma_{\eta\theta\beta}+\gamma_{\eta\theta\beta})
-\Gamma_{\alpha\nu\delta}\bar{G}_{\nu\eta}\bar{G}_{\delta\theta}
\Gamma_{\eta\theta\beta} \nonumber \\
&-&\gamma_{\alpha\nu\delta}\bar{G}_{\nu\eta}\bar{G}_{\delta\theta}
\gamma_{\eta\theta\beta}\Big ], \label{memfun-col} \eea

\noindent We look at the properties of $K_{\alpha\beta}^{(2)}$ in
detail in section \ref{sec_oloop}.

\subsection{Single-particle contribution to vertex function}
\label{sec_scdev4}

Besides the collective contribution to the vertex function
$\Gamma_{\alpha\beta}$, we have the single-particle contribution.
Following the discussion in MMS\cite{MMS} we find that the
single-particle contribution $\gamma_{\alpha\beta}$ to the memory
function $\Gamma_{\alpha\beta}$ at the second order is given by

\be \label{memfun-sin}
\gamma_{\alpha\beta}^{(2)}=-\sum_{\nu\delta\eta\theta}
\gamma_{\alpha{\nu\delta}}\widetilde{G}_{\nu\eta} {\cal
G}_{\delta\theta}\gamma_{\eta\theta\beta} ~~.\ee

\noindent  To avoid ambiguity the collective contribution obtained
in eqn. (\ref{memfun-col}) is to be denoted as
$K_{\alpha\beta}^{(2)}$. The vertex function
$\gamma_{\alpha\beta\delta}$ is defined by Eq.(\ref{eq:4.96}) and
two-point single-particle propagator ${\cal G}$ is defined in
(\ref{eq:54}). The quantity $\widetilde{G}_{\alpha\beta}$ is another
dressed propagator similar to $\bar{G}_{\alpha\beta}$ defined in
eqn. (\ref{bgdef}) corresponding to the collective contribution
$K_{\alpha\beta}$. $\widetilde{G}_{\alpha\beta}$ is defined as

\be \widetilde{G}_{\alpha\beta}=\sum_{\nu\delta\eta\pi}{\cal
G}_{\alpha\nu}\sigma_{\nu\delta}G_{\delta\eta} \sigma_{\eta\pi}
{\cal G}_{\pi\beta} \ee

\noindent In Appendix \ref{appC} we show that like the set
\{$G^{(0)}$,$G$, ${\cal G}$,$\bar{G}$\}, the matrix  $\widetilde{G}$
also satisfies the property of being a FDRMP.

\setcounter{equation}{0}
\section{One-Loop Self-Energy}
\label{sec_oloop}

\subsection{General Structure of One-loop Self-energies}
\label{sec_oloop1}

Generally the self energy matrix $\Sigma$ is defined in terms of the
Dyson equation involving  inverse of the matrix of the two-point
function as follows

\be \label{dyson} \Gamma \equiv G^{-1} = G_0^{-1}-\Sigma ~~. \ee

\noindent In the above equation $G_0^{-1}$ refers to the zeroth
order vertex functions. In field theoretic models this zeroth order
theory stands for the gaussian level theories which usually refers
to the trivial case {\it e.g.}, linear dynamics of fluctuations. In
the present context the zeroth order ( in interaction potentials )
theory refers to the non interacting system. The self energy
$\Sigma$ will be therefore defined here by subtracting the zeroth
order and the first order contributions from the full vertex
function

\be \label{dyson1} -\Sigma = \Gamma - \Gamma^{(0)}-\Gamma^{(1)}
~~~.\ee

At the one-loop level the self energy is a sum of collective and
single particle contributions respectively denoted by $K$ and
$\gamma$,

\be \label{se-2} -\Sigma =\gamma^{(2)}+K^{(2)} + {\cal O}(3)~~,\ee

\noindent  where the superscript $(2)$ indicates the second order in
in the interaction potential and ${\cal O}(3)$ stands for higher
order contributions. The single particle and the collective
contributions are respectively expressed as,

\bea  \label{ogam_s} \gamma^{(2)} &=&
-\frac{\gamma}{2}\widetilde{G}{\cal
G}\gamma=-\frac{1}{4}\gamma\left[(\widetilde{G}+{\cal G})({\cal
G}+\widetilde{G}) -\widetilde{G}\widetilde{G}-{\cal G}{\cal
G}\right]\gamma \\
\label{ogam_c} K^{(2)} &=&
-\frac{1}{4}\left(\gamma\bar{G}\bar{G}\Gamma
+\Gamma\bar{G}\bar{G}\gamma \right ) \nonumber  \\
&=& -\frac{1}{4}(\gamma +\Gamma )\bar{G}\bar{G}(\gamma +\Gamma)
+\frac{1}{4}\gamma \bar{G}\bar{G}\gamma +\frac{1}{4}\Gamma
\bar{G}\bar{G}\Gamma~~~. \nonumber \\ \eea

\noindent  In order to further analyze the one-loop results for
$\gamma$ and $K$ obtained above we introduce the generalized
functional $\Pi[E,A]$ of a two-point propagator $E$ and three-point
vertex $A$ as

\be \Pi [ A,E ]=-\frac{1}{2}E A A E
\label{eq:342}~~.
\ee

\noindent The one-loop contributions to the different elements of
the self energy $\Sigma$ matrix defined in eqn. (\ref{se-2}) are
 now expressed in terms of those of matrix $\Pi$.
We focus on the $\alpha\mu$-th element ($\alpha\mu{\in}\{\rho,B\}$ )
of the matrix $\Pi$:

\bea \Pi_{\alpha\mu} \left [ A,E \right ](-1) &=& \label{gamEA}
 -\sum_{\kappa ,\nu, \sigma ,\delta }\frac{1}{2}
\int d2d3 E^{*}_{\alpha\kappa\nu}(123)\delta (1+2+3) A_{\kappa\sigma
}(2) A_{\nu\delta}(3)E_{\mu\sigma\delta}(123) \nonumber \\
\eea

\noindent where $A$ is a FDRMP and $E$ is a three-point vertex that
satisfies the FDR of section \ref{sec_fdt}. In the above definition
for $\Pi$, the propagator $A$ is from the set $\{G^{(0)},{\cal
G},G,\bar{G},\widetilde{G}\}$ and the three-point vertices $E$
include $\{\gamma^{(0)},\gamma, \Gamma\}$. In terms of the
functional $\Pi[A,E]$, the single-particle and collective
contributions respectively given by eqns. (\ref{ogam_s}) and
(\ref{ogam_c}) are obtained in the form:

\bea \gamma^{(2)} &=& \frac{1}{2}\Pi \left [({\cal
G}+\widetilde{G}),\gamma \right ] -\frac{1}{2}\Pi \left [ {\cal
G},\gamma \right ] -\frac{1}{2}\Pi \left [ \widetilde{G},\gamma
\right ] ~~~, \\
 K^{(2)} &=&  \frac{1}{2}\Pi \left [ \bar{G};\gamma +\Gamma \right ]
-\frac{1}{2}\Pi \left [ \bar{G};\gamma \right ] -\frac{1}{2}\Pi
\left [ \bar{G};\Gamma \right ] ~~~. \eea

\noindent Next we consider the fluctuation-dissipation symmetry of
the single particle and collective contributions to the self enegy
$\Sigma$ at one loop level.

\subsection{Partial Self-energies and the FDT}
\label{sec_oloop2}

We investigate the existence of a fluctuation-dissipation theorem
(FDT) satisfied by the partial self-energy in Fourier space. We
first separate $\Pi [ A,E ]$ into its components, the nonzero
choices for the indices $\kappa,\nu,\sigma,$ and $\delta$. There are
nine contributions to $\Pi_{BB}[A,E]$, five contributions to
$\Pi_{\rho B}[A,E]$ and five contributions to $\Pi_{B\rho}[A,E]$. In
order to better organize the algebra we replace all $A_{\rho\rho}$
internal lines using the FDR:

\be A_{\rho\rho}(q,\omega )=\frac{A_{B\rho}(q,\omega ) -A_{{\rho}
B}(q,\omega )}{i\beta\omega} ~~~. \ee

\noindent For treating the two-point vertices we introduce a set of
simplifying relations which are generalizations of eqns.
(\ref{sreln1})-(\ref{sreln3}) in terms of the generalized vertex
function  $E$ :

\bea \label{compf1}
\phi_i &=& i\beta\omega_{i}E_{B_i\rho_j\rho_k},\\
\label{compf2}\psi_{i} &=& \beta^2\omega_{j}\omega_{k}E_{\rho_i B_jB_k},\\
\label{compf3} P&=&
-i\beta^3\omega_{i}\omega_{j}\omega_{k}E_{B_iB_jB_k}~~. \eea

\noindent where we have explicitly indicated in the first term on
LHS the factor of $\beta$, which was earlier absorbed in the
definition of the frequency $\omega$. The quantities $\phi_i$,
$\psi_i$ and $P$ defined above in terms of the generalized vertex
function $E\in\{\Gamma,\gamma,\gamma^{(0)}\}$ satisfies the same set
of identities given by eqns. (\ref{viden1}) and (\ref{viden2}).
After considerable algebra we find

\bea &&i\omega_{1}\Pi_{BB}[A,E]+\Pi_{\rho B}[A,E]
-\Pi_{B\rho}[A,E] \nonumber \\
&=& -\frac{i}{D}\int d2d3 \Bigg [  \Big \{ A_{B\rho}(2)A_{B\rho}(3)
C_1^*+A_{\rho B}(2)A_{\rho B}(3) C_1 \Big \} \nonumber \\
&-&  \Big \{ A_{B\rho}(2)A_{\rho B}(3) C_2^*+ A_{\rho
B}(2)A_{B\rho}(3)C_2 \Big \} \Bigg ] \delta(1+2+3)~~.
\label{eq:71a}
\eea

\noindent We have defined the quantities $C_1$, $C_2$, and $D$
respectively as,

\bea \label{sreln11}
 C_1 &=& \phi_1^*(\phi_1 +\phi_2 +\phi_3 +\psi_1
+\psi_2+\psi_3+P) \\
\label{sreln12}
C_2 &=&(\phi_1+\phi_2+\psi_3)(\phi_1^*+\psi_2^*+\phi_3^*)-\phi_2\phi_3^* \\
\label{sreln13} D &=& \beta^3 \omega_1\omega_2\omega_3~~. \eea

\noindent Using the identities (\ref{viden1}) and (\ref{viden2}) it
is straightforward to show that $C_1$ and $C_2$ both vanishes,
obtaining a relation similar to (\ref{2p-disp}) to hold with the
elements of the generalized matrix $\Pi$ :

\be \label{fdr-Sigma} \tilde{\cal F}[\Pi]\equiv
-i\beta\omega_1\Pi_{BB}(1)+\Pi_{B\rho}(1)-\Pi_{\rho B}(1)=0~~~~.\ee

\noindent  In the above equation and in what follows the functional
dependence of $\Pi[E,A]$ on the vertex $E$ and correlation $A$ are
not explicitly shown in the RHS to avoid cluttering.  In Section
\ref{sec_fdt} the full two-point vertex funtion matrix $\Gamma_{ij}$
was shown to be a FDRMP with the result (\ref{twovf-fdt}). We have
demonstrated here that the self energy matrix
$\Pi_{\alpha\beta}[E,A]$ expressed in terms of the generalized
vertex $E_{\alpha\beta\delta}$ and correlation functions
$A_{\beta\delta}$, are also FDRMP. Since $\tilde{\cal F}$ defined in
eqn. (\ref{fdr-func}) is linear, using eqn. (\ref{fdr-Sigma}) in the
definitions (\ref{ogam_c}) and (\ref{ogam_s}) respectively, it
follows that

\bea \label{fdr-Kcol} \tilde{\cal F}\left [K^{(2)} \right ]=0~~, \\
\label{fdr-Ksin} \tilde{\cal F}\left [\gamma^{(2)}\right ]=0~~. \eea

\noindent Hence both the collective and single-particle one loop
level contributions to the self energy respectively denoted by
$K_{ij}$ and $\gamma_{ij}$ are FDRMP.

After making full use of the vertex identities we find that the self
energy matrix element $\Pi_{B\rho}(-1)$ is the sum of two pieces

\be -2\Pi_{B\rho}(-1)= i\int \frac{d2 d3}{D} \delta(1+2+3) \Bigg [
{(\phi_1^*)}^2 A_{\rho B}(2)A_{\rho B}(3) + {(\phi_2)}^2 A_{\rho
B}(2)A_{B\rho }(3) \Bigg ] ~~~. \ee

\noindent Notice that this self-energy depends only on the
three-point vertex functions with one $B$ label; $E_{B\rho\rho}$ and
$E_{\rho B\rho}$ respectively as follows from eqn. (\ref{compf1}).

\subsubsection{Separation into High and Low Frequency Components}
\label{sec_oloop3}

We will be interested in low-frequency-long-time phenomena. We
identify the low-frequency contribution to the vertex using the
Vertex Theorem (see appendix \ref{appB}) as :

\be {E}_{\rho B \rho}(0,0,0;q_1,k_2,-q_1-k_2) =
\gamma_{\rho\rho\rho}(q_1,k_2,-q_1-k_2) \ee

\noindent where $\gamma_{\rho\rho\rho}$ is the static 3-point vertex
and write the frequency dependent vertex functions as,

\bea {E}_{\rho B \rho}(\omega_1,\omega_2,\omega_3;q_1,k_2,-q_1-k_2)
&=& \gamma_{\rho\rho\rho}(q_1,k_2,-q_1-k_2) \nonumber \\
&+& \Delta_{\rho B
\rho}(\omega_1,\omega_2,\omega_3;q_1,k_2,-q_1-k_2)~~. \eea

\noindent The one loop contribution denoted by  $\Pi[E,A]$ is split
in to a low frequency part $\widehat{\Pi}$ (for
$\omega{\rightarrow}0)$ and a high frequency part $\widetilde{\Pi}$

\be \label{bartild} \Pi_{B\rho}(1)= \widehat{\Pi}_{B\rho}(1)
+\widetilde{\Pi}_{B\rho}(1) \ee

\noindent where $\widehat{\Pi}_{B\rho}(1)$ is the contribution with
${E}_{\rho B \rho}$ replaced by its static limit value. After a
large amount of algebra we find the  low frequency contribution
$\widehat{\Pi}(1)$ is  reduced to the form,

\bea \label{barSig} \widehat{\Pi}_{B\rho} (1) &=&
\frac{1}{2\beta}\hat{\cal O} \bar{A}_{\rho\rho}(k_2)
\bar{A}_{\rho\rho}(k_3)
{[\gamma_{\rho\rho\rho}(q_{1},k_{2},k_{3})]}^2 \\
&+& \frac{\omega_1}{2\beta}\hat{\cal O} \int\frac{dx}{2\pi}
\int\frac{dy}{2\pi} {[\gamma_{\rho\rho\rho}(q_1,k_2,k_3)]}^2
\frac{A_{\rho\rho}(x,k_2)A_{\rho\rho}(y,k_3)}
{-\omega_{1}+x+y-i\eta}~~, \nonumber \eea

\noindent where we have defined the operator $\hat{\cal O}$ as

\be \hat{\cal O}=\int\frac{d^{d}k_{2}}{(2\pi)^{d}}
\int\frac{d^{d}k_{3}}{(2\pi)^{d}} \delta (q_{1}-k_{2}-k_{3}) ~~~.
\ee

\noindent The static ($\omega\rightarrow{0}$ ) contribution
$\widehat{\Pi}(1)$ is given by

\be \widehat{\Pi}_{B\rho}(q_1,0) = \frac{1}{2\beta}\hat{\cal O}
\bar{A}_{\rho\rho}(k_{2}) \bar{A}_{\rho\rho}(k_3)
{[\gamma_{\rho\rho\rho}(q_1,k_2,k_3)]}^2 ~~. \label{SigBr0} \ee

\noindent Since the low-frequency part $\widehat{\Pi}$ is itself a
FDRMP, the $BB$ element of the $\widehat{\Pi}$ matrix is given by

\bea \label{SigBB1} \widehat{\Pi}_{BB}(1) &=&
-\frac{2\beta^{-1}}{\omega_{1}} \mathrm{Im} \left [
\widehat{\Pi}_{B\rho}(1)
\right ] \nonumber \\
&=& -\frac{1}{2\beta^2} \hat{\cal O} \int\frac{dx}{2\pi}
{[\gamma_{\rho\rho\rho}(q_1,k_2,k_3)]}^2
A_{\rho\rho}(x,k_2)A_{\rho\rho}(\omega_1-x,k_3) ~~~. \eea

\noindent Notice that it follows directly from eqns. (\ref{barSig})
and (\ref{SigBB1}) that the sum rule

\be \label{FDreln1} \widehat{\Pi}_{B\rho}(q,0)
=-\int\frac{d\omega}{2\pi}\beta\widehat{\Pi}_{BB}(q,\omega ) \ee

\noindent is satisfied. The high-frequency contribution
$\widetilde{\Pi}_{B\rho}(\omega)$ to the self energy  satisfy

\be \frac{\widetilde{\Pi}_{B\rho}(1)}{\widehat{\Pi}_{B\rho}(1)}
\approx \omega_{1} \ee as $\omega_{1}\rightarrow 0$,  and hence does
not contribute to the slow dynamics of the system. A key result in
this work is the theorem that $\Gamma_{B\rho\rho}$ reduces to the
static three-point vertex in the low-frequency limit.

\subsection{Original One loop Problem}
\label{sec_oloop4}

We now focus  on the one loop contributions for collective and
single-particle self-energies respectively denoted by
$K_{\alpha\mu}$ and $\gamma_{\alpha\mu}$ which were introduced
earlier in section \ref{sec_oloop}. For the collective contribution
$K$ the appropriate three-point vertex functions are
$E\equiv\{(\gamma+\Gamma),\gamma,\Gamma\}$, while the correlation
function $A\equiv\bar{G}$. The collective contribution is divided in
to two parts as

\be K_{\alpha\mu}(1)=\widehat{K}_{\alpha\mu}(1) +
\widetilde{K}_{\alpha\mu}(1) \ee

\noindent  where  we have followed the same notations in terms of
bar and tilde on the respective terms as given in eqn.
(\ref{bartild}) in indicating the low and high frequency components.
Since $K_{\alpha\mu}$ is a FDRMP, we obtain from eqn. (\ref{ogam_c})
and (\ref{SigBB1}), the low frequency part of the collective
contribution $K^{(2)}_{BB}$ as

\bea \widehat{K}_{BB}^{(2)}(1) &=& -\frac{2\beta^{-1}}{\omega_{1}}
\mathrm{Im} \widehat{K}^{(2)}_{B\rho}(1) \nonumber \\
&=& -\frac{1}{\beta^2}\hat{\cal O}
\int\frac{dx}{2\pi}\bar{G}_{\rho\rho}(x,k_{2}) \int\frac{dy}{2\pi}
\bar{G}_{\rho\rho}(y,k_{3}) \pi\delta (\omega_{1}-x-y) \nonumber \\
&\times&
\left[\frac{1}{2}{\{(\gamma_{\rho\rho\rho}+\Gamma_{\rho\rho\rho})\}}^{2}
-\frac{1}{2}(\gamma_{\rho\rho\rho})^{2}
-\frac{1}{2}(\Gamma_{\rho\rho\rho})^{2}\right] \nonumber \\
&=&-\frac{1}{2\beta^2}\hat{\cal O} \int\frac{dx}{2\pi}
\bar{G}_{\rho\rho}(x,k_{2})\bar{G}_{\rho\rho}(\omega_1-x,k_{3})
\gamma_{\rho\rho\rho}\Gamma_{\rho\rho\rho}~~. \label{KBB2C}\eea

\noindent For the single particle contribution $\gamma_{\alpha\mu}$
the appropriate three-point vertex functions are $E\equiv\gamma$,
while the correlation function $A\equiv\{({\cal
G}+\widetilde{G}),{\cal G},\widetilde{G}\}$. We obtain from eqn.
(\ref{ogam_s}) the single particle contribution as,

\be \widehat{\gamma}_{BB}^{(2)}(1) =-\frac{1}{2\beta^2}\hat{\cal O}
\int\frac{dx}{2\pi} \widetilde{G}_{\rho\rho}(x,k_{2}) {\cal
G}_{\rho\rho}(\omega_1-x,k_{3})\gamma^{2}_{\rho\rho\rho} ~~.
\label{KBB2S}\ee

\noindent Eqns. (\ref{KBB2C}) and (\ref{KBB2S}) are approximations
for the second-order collective and single particle contributions to
the frequency dependent self energy $\Sigma$ {\bf including} vertex
corrections. We write for the respective collective and single
particle contributions in the time regime

\bea \widehat{K}_{BB}^{(2)}(q_{1},t_{1})
&=&-\frac{1}{2\beta^2}\hat{\cal O}
\bar{G}_{\rho\rho}(t_{1},k_{2})\bar{G}_{\rho\rho}(t_{1},k_{3})
\gamma_{\rho\rho\rho}\Gamma_{\rho\rho\rho}~~. \label{KBB2C1}\eea

\noindent and

\be \widehat{\gamma}_{BB}^{(2)}(q_{1},t_{1})
=-\frac{1}{2\beta^2}\hat{\cal O}
\widetilde{G}_{\rho\rho}(t_{1},k_{2}) {\cal
G}_{\rho\rho}(t_{1},k_{3})\gamma^{2}_{\rho\rho\rho} ~~.
\label{KBB2S1}\ee

\noindent In the low-frequency limit we assume, as we show
self-consistently, $\widehat{K}_{BB}^{(2)}(q_{1},z_{1})$ grows
arbitrarily large as the frequency $z_1$  goes to zero.  Since
$\widehat{\gamma}_{BB}^{(2)}(q_{1},0)$ is regular we can drop the
single-particle contribution to the memory function in the kinetic
equation.

In the expression  given in Eq. (\ref{KBB2C1}) for the memory
function, the implication of having the full vertex function
$\Gamma_{\rho\rho\rho}$ on the dynamic behavior of the fluid has
been ignored. To lowest order in the interaction potential, the
vertex $\Gamma_{\rho\rho\rho}$ is approximated by the corresponding
quantity $\gamma_{\rho\rho\rho}$ of the non-interacting theory which
is the case for the ideal gas. Since the vertex functions are only
involved here ( see Appendix B) in the $\omega{\rightarrow}0$ limit
and hence they appears like a static quantity. So far these static
approximations are not directly connected with HNC or PY
approximations. We discuss the role of vertex corrections elsewhere.
We have then

\be \Gamma_{\rho\rho\rho}= -\gamma_{\rho\rho\rho}
=-\frac{1}{\bar{\rho}^{2}} ~~~. \ee

\noindent We also assume consistent with the long-time approximation
that for low-frequencies

\be G_{\rho\rho}(q,\omega )\gg G_{\rho B}(q,\omega) \ee

\noindent and in $\bar{G}_{\rho\rho}(q, \omega )$ the
$G_{\rho\rho}(q, \omega )$ term dominates and its coefficient can be
replaced by its $\omega =0$ value.  Using the definitions
$G_{\rho\rho}(q, \omega)=\bar{\rho}S(q)F(q,\omega )$  we find

\bea {\bar{G}_{\rho\rho}(q, \omega)} &=& \beta
V(q)\bar{\rho}G_{\rho\rho}(q, \omega) \nonumber \\
&=& \tilde{V}(q)\bar{\rho}S(q)F(q, \omega)~~, \eea

\noindent where we have denoted $\tilde{V}=\bar{\rho}\beta V(q)$ as
the scaled potential. From the above definitions we obtain  the
relation $\tilde{F}(q,t)=\tilde{V}(q)S(q)F(q,t)$.

\noindent We can then write in the long time regime the simple
result

\be
K_{BB}^{(2)}(q,t) = -\frac{1}{2\bar{\rho}^2\beta^2} \int
\frac{d^{d}k}{(2\pi )^{d}} \tilde{F}(q-k,t)\tilde{F}(k,t)
\label{FR}
\ee

We need the Fourier-Laplace transform of Eq.(\ref{nep-integ}) to go
back in the low-frequency form of the kinetic equation. After some
simple rearrangements we obtain our main result  from Eq.(\ref{eq:4.35})

\be \frac{F(q,z)}{1-zF(q,z)}= -i\frac{S(q)}{2\bar{\rho}}
\int_{0}^{\infty}dt e^{izt}
\int
\frac{d^{d}k}{(2\pi )^{d}} \tilde{F}(q-k,t)\tilde{F}(k,t)
\label{eq:FRE} \ee

\noindent This is a highly nonlinear equation for $F(q,t)$.  This
result is identical, after reconciling notation, to the result in
Ref. \cite{SDENE} for Smoluchowsky dynamics. As in SM we can carry
out an analytic treatment of Eq.(\ref{nep-integ}) by expanding

\be F(q,z)=\frac{f(q)}{z}+\psi (q,z) \ee

\noindent for $z\rightarrow 0$ and $\lim_{z\rightarrow 0}z \psi
(q,z)\rightarrow 0$ while $\lim_{z\rightarrow 0} \psi
(q,z)\rightarrow \infty$. $f(q)$ is the nonergodicity parameter with
the interpretation

\be \lim_{t\rightarrow\infty}G_{\rho\rho}(q,t)=f(q)\bar{\rho}S(q)
~~~. \ee

\noindent In the small $z$ or long time limit the eqn.
(\ref{eq:FRE}) reduces to an intergal equation in terms of the
nonergodicity parameters $f(q)$ of the form

\bea \frac{f(q)}{1-f(q)}= \frac{S(q)}{2\bar{\rho}} \int
\frac{d^{d}k}{(2\pi )^{d}} S(q-k)
\tilde{V}(q-k)f(q-k)S(k)\tilde{V}(k)f(k) \nonumber \\
\label{eq:FREA}
~~~.
\eea

This equation must be supplemented with an equation connecting
the potential and the static structure factor.  At second order in the
potential we have

\bea  S^{-1}(q)-1=\tilde{V}(q)-\frac{1}{2\bar{\rho}}
\int\frac{d^{d}k}{(2\pi )^{d}}\tilde{V}(q-k)S(q-k)\tilde{V}(k)S(k)
\nonumber \\
\label{inteqn-V} \eea

We choose to fix $S(q)$ and solve for the pseudo-potential
$\tilde{V}(q)$.
The question of a pseudo-potential is discussed in detail in SDENE.
We can determine $\tilde{V}(q)$
once we have a form for the static structure factor.
Typically we have used the exact solution of the approximate
Percus-Yevick\cite{PY} equation for hard spheres.  One can then
carry our the explicit determination of the quantities
characteristic of the ENE transition.

The above integral equations for $f(q)$ is
solved over a grid to obtain the nonzero solutions using iterative
methods. The density at which the trivial solution $f(q)=0$ changes
to a nonzero set of nonergodicity parameter values marks the
location of the ideal ENE transition in the dense liquid approaching
from the liquid side.

In Fig. \ref{cfig1} we display the static structure factor for the
hard sphere liquid as obtained from Percus-Yevick\cite{PY} equations
with Verlet-Weiss corrections \cite{V-W} at packing fraction $\eta
=0.62$. Using this $S(k)$ as an input we solve iteratively eqns.
(\ref{inteqn-V}) to obtain the renormalized potential
$\tilde{V}(q)$. The wave vector $q$ is chosen over a grid of upper
cutoff value $q\sigma=80$ and having 500 points. Fig. \ref{cfig2}
displays the $\tilde{V}(q)$ obtained with input structure factor of
Fig. \ref{cfig1}.

Next we use this renormalized potential to evaluate the vertex
functions in eqn. (\ref{eq:FREA}) for the the nonergodicity
parameters $f(q)$. Solving these integral equations iteratively we
obtain that the $f(q)$ are vanishing till the critical packing
fraction of $\eta=.62$. The corresponding nonergodicity parameters
$f(q)$ over the whole wave vector grid is shown in Fig. \ref{cfig3}.
The ENE transition point of the core problem of self-consistent feed
back mechanism using Percus-Yevick strucutre factors with
Verlet-Weiss corrections therefore at packing fraction of $.62$
which is close to the close-packing density.

\setcounter{equation}{0}
\section{Conclusions}
\label{sec_con}

We have shown within the pseudo-potential expansion the ENE problem
for ND reduces to precisely the same problem for SD. For
hard-spheres and using the PY approximation for the structure factor
we can solve the approximate ENE problem to find $\eta^{*}=0.62$, and the
two-step exponents $a$ and $b$ consistent with the results found in
SM.  This is a nontrivial result and is closely associated with the vertex
theorem which relates the low-frequency limit of
$\Gamma_{B\rho\rho}$ to the static vertex $\gamma_{\rho\rho\rho}$.
While we have proved this result only to second order in perturbation
theory, we expect it is more general\cite{GLD}.

We have found simplification of the ND case only in the low-frquency
regime. More generally the SD case is considerably simpler as
demonstrated in MMS where the zeroth order vertices are very simple.

This result for the dynamic structure factor raises questions about
higher-order correlation functions such as $G_{\rho\rho\rho}(123)$.
There is good reason to believe that we can make progress in determining
this quantity in the low-frequency regime.

Our results here depend on ignoring static-vertex corrections and
higher-order loop contributions.  We need to investigate these
correction to see if the results for the ENE are stable.  We see
from Eq.(\ref{KBB2C1}) that the low-frequency dynamics depends not
just on the static structure factor, but also on the static
three-point vertex.


The theory  presented in this work has been set up to deal with what
we call the {\it core} problem in the ND case. The primary
motivation here is the determination of the observables involving
these core variables $\Phi_{0}=(\rho , B)$. The terms core  refers
to the fact that the Hamiltonian and hence the corresponding MSR
action is expressed in terms of only these variables.  As already
pointed out ( see sec. 9 in ND1), by going beyond the core
variables, considerations of additional degrees of freedom enter the
theory. For example in the ND case, we have a larger phase-space due
to the momentum degrees of freedom and there are additional
conservation laws for the system. If one extends the set of core
variables to include the momentum density, we have all the
correlation functions among $\rho$, $B$ and $g$ in the description.
The resulting hierarchical structure in the formal expressions for
the correlation functions may change the viability of the ENE
transition being discussed here. In this regard it is useful to note
the known results of including the momentum density fluctuations in
the fluctuating hydrodynamic approach to the problem. The momentum
density coupling to the density fluctuations in the hydrodynamic
equations give rise to the ergodicity restoring mechanisms
\cite{dm,dm09}. In the case of SD on the other hand, the core
problem itself covers essentially all of the degrees of freedom of
interest. However the microscopic dynamics in this case is
dissipative and the equations of motion for the collective
collective variables \cite{dean,kawasaki} involve multiplicative
noise which may have implications in restoring ergodic behavior in a
liquid. The present work demonstrates that in the case of the core
problem through a reorganization of the perturbation theory in terms
of an effective potential, the ideal ENE transition is pushed to the
close pack density for a hard sphere system.


\newpage
\renewcommand{\thesection}{\Alph{section}}
\renewcommand{\theequation}{\Alph{section}\arabic{equation}}
{\appendix
\section{Three point vertex functions}
\setcounter{equation}{0}
\label{appA}

We obtain the FDT relations between the different three-point vertex
functions here. For this we use the general definition of the three
point correlation functions and the corresponding three-point vertex
function as,

\be \label{AG3v3} G_{\alpha_i{\mu_j}\nu_k} =
-G_{\alpha\alpha^\prime}(i)G_{\mu\mu^\prime}(j)G_{\nu\nu^\prime}(k)
\Gamma_{{\alpha_{i}^\prime}{\mu_{j}^\prime}{\nu_{k}^\prime}}\ee

\noindent where $\alpha_i$ stands for the field $\alpha$ at the
point $i$ and so on where everything is in Fourier space.
The labels $\alpha$,$\mu$ and $\nu$ are
respectively taken from the set $\{\rho,B\}$. Using the result that
$\Gamma_{\rho_i\rho_j\rho_k}=0$, we obtain from the formula
(\ref{AG3v3}), the various cumulants are :

\bea \label{AG3e1} -G_{\rho_i\rho_j\rho_k} &=& G_{\rho\rho
}(i)G_{\rho \rho}(j)G_{\rho B}(k) \Gamma_{\rho_i \rho_jB_k} +
G_{\rho\rho }(i)G_{\rho
B}(j)G_{\rho\rho}(k)\Gamma_{\rho_i B_j\rho_k } \nonumber \\
&+& G_{\rho B}(i)G_{\rho \rho}(j)G_{\rho\rho}(k)\Gamma_{B_i\rho_j
\rho_k} + G_{\rho \rho}(i)G_{\rho B}(j)G_{\rho B}(k)\Gamma_{\rho_i
B_j B_k } \nonumber \\
&+& G_{\rho B}(i)G_{\rho\rho}(j)G_{\rho B}(k)\Gamma_{B_i\rho_jB_k} +
G_{\rho B}(i)G_{\rho B}(j)G_{\rho\rho}(k)\Gamma_{B_iB_j\rho_k}
\nonumber \\
&+& G_{\rho B}(i)G_{\rho B}(j)G_{\rho B}(k)\Gamma_{B_iB_jB_k} \\
\label{AG3e2}-G_{B_i\rho_j \rho_k} &=& G_{B\rho }(i) G_{\rho
B}(j)G_{\rho\rho}(k)\Gamma_{\rho_i B_j\rho_k} + G_{B\rho
}(i)G_{\rho\rho}(j)G_{\rho B}(k)
\Gamma_{\rho_i\rho_j B_k} \nonumber \\
&+& G_{B\rho}(i)G_{\rho B}(j)G_{\rho B}(k)\Gamma_{\rho_iB_jB_k} \\
\label{AG3e3} - G_{B_iB_j\rho_k} &=& G_{B\rho}(i)
G_{B\rho}(j)G_{\rho{B}}(k)\Gamma_{\rho_i \rho_j B_k} ~~~.\eea

\noindent Substituting the results (\ref{AG3e1})-(\ref{AG3e3}) in
eqn. (\ref{G3e0}), we obtain an expansion in terms of the various
three-point vertex functions. For example the coefficient of the
vertex function $\Gamma_{B_i\rho_j\rho_k}$ is obtained after some
trivial but tedious algebra as

\bea \label{G3e6} \Gamma_{B_i\rho_j\rho_k}  &:& i\omega_i
G_{\rho{B}}(i)G_{B\rho}(j) G_{B\rho}(k) + {\omega_i\omega_j} G_{\rho
B}(i)G_{\rho\rho}(j)G_{B\rho
}(k) \nonumber \\
&=& i\omega_i G_{\rho{B}}(i) G_{\rho{B}}(j) \Big \{ G_{B\rho}(k)-
i\omega_k G_{\rho\rho}(k)\Big \}=i\omega_i G_{\rho{B}}(i)
G_{\rho{B}}(j)G_{\rho{B}}(k) \eea

\noindent In reaching the above result we have used the FDT relation
(\ref{twovf-fdt}).  Similarly for the other two vertices with one
$B$ field, {\em i.e.}, $\Gamma_{\rho_iB_j\rho_k}$ and
$\Gamma_{\rho_i\rho_jB_k}$ are obtained. For the vertex with two $B$
fields, {\em e.g}, $\Gamma_{B_iB_j\rho_k}$  we obtain the
corresponding coefficient as

\bea \label{G3e9} \Gamma_{B_iB_j\rho_k}  &:&
{\omega_i\omega_j}G_{\rho B}(i)G_{\rho B}(j)G_{B\rho}(k) -
i\omega_i\omega_j\omega_k G_{\rho B}(i)G_{\rho B}(j)G_{\rho\rho}(k)
\nonumber \\
&=&{\omega_i\omega_j}G_{\rho B}(i)G_{\rho B}(j)G_{\rho{B}}(k) \eea

\noindent and similarly we obtain the coefficients of the other two
vertices each with two $B$ fields. After organizing the coefficients
of the different three-point vertex functions the following result
is obtained.

\bea \label{Avcond1} && i\Big ( \omega_i\Gamma_{B_i\rho_j\rho_k}+
\omega_j \Gamma_{\rho_iB_j\rho_k} + \omega_k
\Gamma_{\rho_i\rho_jB_k} \Big ) -
i \omega_i\omega_j\omega_k\Gamma_{B_iB_jB_k} \nonumber \\
&+& \Big \{ {\omega_i\omega_j}\Gamma_{B_iB_j\rho_k}+
{\omega_j\omega_k}\Gamma_{\rho_iB_jB_k} +
{\omega_k\omega_i}\Gamma_{B_i\rho_jB_k} \Big \}=0 \eea

\noindent In reaching the above result we have dropped a nonzero
common factor of $G_{\rho{B}}(i)G_{\rho{B}}(j)G_{\rho{B}}(k)$ from
the LHS.

Next, consider the relations (\ref{eBBr}) involving three-point
cumulants having {\em two} $B$ fields. Once again substituting the
results (\ref{AG3e2})-(\ref{AG3e3}) in eqn. (\ref{eBBr}), we obtain
after organizing the coefficients of the different three-point
vertex functions the following result.

\bea \label{AeBBr8} && G_{B\rho}(k)\Big \{
{i\omega_k}\Gamma^*_{\rho_i \rho_j B_k }+ i\omega_i
\Gamma_{B_i\rho_j\rho_k} +i\omega_j \Gamma_{\rho_iB_j\rho_k} +
\omega_i\omega_j \Gamma_{B_iB_j\rho_k}
\Big \} \nonumber \\
&=&G_{\rho{B}}(k) \Big \{ i\omega_i \Gamma_{B_i\rho_j\rho_k}+
i\omega_j \Gamma_{\rho_iB_j\rho_k} + {i\omega_k}\Gamma_{\rho_i
\rho_j B_k} \nonumber \\
&+& \omega_i\omega_j \Gamma_{B_iB_j\rho_k} +\omega_i\omega_k
\Gamma_{B_i\rho_jB_k}  + \omega_j\omega_k \Gamma_{\rho_iB_j B_k } -
i\omega_i\omega_j \omega_k \Gamma_{B_iB_jB_k} \Big \} =0 ~~~. \eea

\noindent In obtaining the above result, we have dropped the common
factor of $G_{\rho{B}}(i) G_{\rho{B}}(j)$ from both sides and used
the basic FD relation (\ref{twovf-fdt}). Using eqn. (\ref{vcond1})
in the result (\ref{AeBBr8}) we find that the coefficient of
$G_{\rho B}(k)$vanishes and one has the result

\be \label{AeBBr9} {i\omega_k}\Gamma^*_{\rho_i \rho_j B_k }+
i\omega_i \Gamma_{B_i\rho_j\rho_k} +i\omega_j
\Gamma_{\rho_iB_j\rho_k} + \omega_i\omega_j \Gamma_{B_iB_j\rho_k} =0
~~~.
\ee

\setcounter{equation}{0}
\section{Higher-Order Thermodynamic Sum Rule}
\label{appB}

\subsection{Three-point Quantities}

Let us consider the low frequency behavior of the full three-point
cumulant $G_{\rho BB}(123)$. The quantity $\rho (1)$ in $G_{\rho
BB}(123)$ can be replaced by an arbitrary function of density as
long as each density corresponds to the same time. The three-point
vertex  $\Gamma_{B\rho\rho}(123)$ is related to the 3-point
correlation $G_{\rho{BB}}$ by the general relation

\be \Gamma_{B\rho\rho}(123)=-\Gamma_{B\rho}(1)\Gamma_{\rho B}(2)
\Gamma_{\rho B}(3) G_{\rho BB}(123) \label{eq:526} ~~~. \ee

\noindent Among the FDR identities for the 3-point cumulants, we
have for the imaginary part of $G_{\rho{BB}}$ the relation :

\be G_{\rho BB}''(123)=\frac{\beta\omega_{2}}{2}G_{\rho\rho B}'(123)
+\frac{\beta\omega_{3}}{2}G_{\rho B\rho }'(123) ~~~. \label{eq:541}
\ee

\noindent Next, look at the definition of the inverse time fourier
transform

\be \label{ft-GrBB} G_{\rho BB}(q;t_{1},t_{2},t_{3})
=\int\frac{d\omega_{1}}{2\pi}
\int\frac{d\omega_{2}}{2\pi}e^{-i\omega_{1}(t_{1}-t_{3})}
e^{-i\omega_{2}(t_{2}-t_{3})}
\widetilde{G}_{\rho{BB}}(\omega_{1},\omega_{2}) \ee

\noindent where we have introduced

\be G_{\alpha\beta\gamma}(123) =
\widetilde{G}_{\alpha\beta\gamma}(\omega_{1},\omega_{2},q_{1},q_{2},q_{3})
\delta (\omega_{1}+\omega_{2}+\omega_{3}) ~~~. \label{eq:543} \ee

\noindent In this notation we do not always write the third
frequency entry since it is implied:

\be \widetilde{G}_{\alpha\beta\gamma}(\omega_{1},\omega_{2})
=\widetilde{G}_{\alpha\beta\gamma}(\omega_{1},\omega_{2},-\omega_{1}-\omega_{2})
\ee

\noindent and the wave number dependence has been suppressed.
Setting $t_{2}=t_{3}$ in eqn. (\ref{ft-GrBB})  we obtain,

\be G_{\rho BB}(t_{1},t_{2},t_{2}) =\int\frac{d\omega_{1}}{2\pi}
\int\frac{d\omega_{2}}{2\pi}e^{-i\omega_{1}(t_{1}-t_{2})}
\widetilde{G}_{\rho BB}(\omega_{1},\omega_{2}) ~~~. \ee

\noindent This vanishes for  $t_{2}>t_{1}$. This is consistent with

\be \int\frac{d\omega_{2}}{2\pi}
\widetilde{G}_{\rho{BB}}(\omega_{1},\omega_{2}) \ee

\noindent being analytic in the UHP for $\omega_{1}$. Assuming
$\widetilde{G}_{\rho BB}(\omega_{1},\omega_{2})$ is analytic in the
UHP for $\omega_{1}$ we can write a dispersion relation

\be \widetilde{G}_{\rho BB}(\omega_{1},\omega_{2})
=\int\frac{d\bar{\omega}}{\pi}\frac{ \widetilde{G}_{\rho
BB}''(\bar{\omega},\omega_{2})} {\bar{\omega}-\omega_{1}-i\eta} ~~~.
\label{eq:546} \ee

\noindent Putting Eq.(\ref{eq:543}) in Eq.(\ref{eq:541}) gives

\be \widetilde{G}_{\rho BB}''(\omega_{1},\omega_{2})
=\frac{\beta\omega_{2}}{2}\widetilde{G}_{\rho\rho
B}'(\omega_{1},\omega_{2}) -\frac{\beta (\omega_{1}+\omega_{2})}{2}
\widetilde{G}_{\rho B\rho }'(\omega_{1},\omega_{2}) \label{eq:547}
~~~. \ee

\noindent Putting Eq.(\ref{eq:547}) in Eq.(\ref{eq:546})  gives

\be \widetilde{G}_{\rho BB}(\omega_{1},\omega_{2})
=\int\frac{d\bar{\omega}}{2\pi}\frac{\beta\omega_{2}
\widetilde{G}_{\rho\rho B}'(q;\bar{\omega},\omega_{2})
-\beta[\bar{\omega}+\omega_{2}] \widetilde{G}_{\rho B\rho
}'(\bar{\omega},\omega_{2})} {\bar{\omega}-\omega_{1}-i\eta} ~~~.
\ee

\noindent Letting $\omega_{1}$ and $\omega_{2}$ go to zero gives

\be \widetilde{G}_{\rho BB}(0,0) =-\beta
\int\frac{d\bar{\omega}}{2\pi} \widetilde{G}_{\rho B\rho
}'(\bar{\omega},0) \label{eq:549} \ee

\noindent We have a FDR identity

\be \widetilde{G}_{\rho B\rho }''(\omega_{1},\omega_{2})
=\frac{1}{2}\beta \omega_{2}\widetilde{G}_{\rho \rho\rho
}(\omega_{1},\omega_{2}) \ee

\noindent which tells us that

\be \widetilde{G}_{\rho B\rho }''(\omega_{1},0)=0 \ee

\noindent and we can write

\be \widetilde{G}_{\rho BB}(q;0,0) =-\beta
\int\frac{d\bar{\omega}}{2\pi} \widetilde{G}_{\rho B\rho
}(\bar{\omega},0) \label{eq:552} ~~~. \ee

\noindent In the time domain

\be G_{\rho B\rho}(t_{1},t_{2},t_{3}) =\int\frac{d\omega_{1}}{2\pi}
\int\frac{d\omega_{2}}{2\pi}e^{-i\omega_{1}(t_{1}-t_{3})}
e^{-i\omega_{2}(t_{2}-t_{3})} \widetilde{G}_{\rho
B\rho}(\omega_{1},\omega_{2}) ~~~. \ee

\noindent Fourier transforming over $t_{2}$ obtains

\be G_{\rho B\rho}(t_{1},\omega_{2},t_{3})
=\int\frac{d\omega_{1}}{2\pi} e^{-i\omega_{1}(t_{1}-t_{3})}
e^{i\omega_{2}t_{3}} \widetilde{G}_{\rho
B\rho}(\omega_{1},\omega_{2},-\omega_{1}-\omega_{2}) ~~~. \ee

\noindent Setting $t_{3}=t_{1}$ and letting $\omega_{2}\rightarrow
0$:

\be G_{\rho B\rho}(t_{1},0,t_{3}) =\int\frac{d\omega_{1}}{2\pi}
\widetilde{G}_{\rho B\rho}(\omega_{1},0,-\omega_{1}) ~~~.
\label{eq:554} \ee

\noindent Combining Eq.(\ref{eq:552}) and (\ref{eq:554}) gives

\be \widetilde{G}_{\rho BB}(0,0,0) =-\beta G_{\rho
B\rho}(t_{1},0,t_{1}) ~~~. \ee

\noindent When the times of the $\rho$'s are equal in
$G_{\rho{B}\rho}$ we have

\be G_{\rho B\rho}(q;t_{1},t_{2},t_{1})
=\int\frac{d\omega_{1}}{2\pi} \int\frac{d\omega_{2}}{2\pi}
e^{-i\omega_{2}(t_{2}-t_{1})} \widetilde{G}_{\rho
B\rho}(\omega_{1},\omega_{2}) \ee

\noindent Introducing $f(t_{1})=\delta \rho (q_{1}t_{1})\delta \rho
(q_{2},t_{1})$ one has $G_{fB}(0)$ satisfies the two-time FDR

\be G_{fB}(0)=-\beta\int\frac{d\bar{\omega}}{2\pi}
G_{f\rho}(\bar{\omega}) =-\beta \langle \delta\rho (q_{3})f\rangle
\ee \be =-\beta S_{\rho\rho\rho}(q_{1},q_{2},q_{3}) \ee

\noindent and the static three-point cumulant enters the
development. We have then

\be \widetilde{G}_{\rho BB}(q;0,0)=-\beta G_{fB}(q;0)
=\beta^{2}S_{\rho\rho\rho}(q_{1},q_{2},q_{3}) ~~~. \ee

\noindent In terms of the three-point vertex

\bea \Gamma_{B\rho\rho}(0,0,0) &=& -\Gamma_{B\rho}(0)\Gamma_{\rho
B}(0) \Gamma_{\rho B}(0) G_{\rho BB}(0,0,0) \nonumber \\
&=& \frac{\beta^{2}S_{\rho\rho\rho}(q_{1},q_{2},q_{3})} {(-\beta
S_{\rho\rho}(q_{1})) (-\beta S_{\rho\rho}(q_{2})) (-\beta
S_{\rho\rho}(q_{3}))} \nonumber \\
&=&-\beta^{-1}\gamma_{\rho\rho\rho}(q_{1},q_{2},q_{3}) ~~~. \eea

\noindent which is a result of much use in evaluating the one-loop
contribution to the self energy both at single-particle and
collective levels. Note that $\gamma_{\rho\rho\rho}$ is a static
three-point vertex.

\setcounter{equation}{0}
\section{FDR Matrix Propagators}
\label{appC}

FDR matrix propagators (FDRMP)
$A_{\mu\nu}(q,\omega )$ satisfy the following
properties:

\bea A_{\mu\nu}(q,\omega ) &=& A_{\nu\mu}^{*}(q,\omega ) \nonumber \\
+i\beta \omega A_{\rho\rho}(q,\omega ) &=& A_{B\rho}(q,\omega
)-A_{\rho{B}}(q,\omega ) \nonumber \\
A_{\rho B}(q,\omega ) &=& \int\frac{d\bar{\omega}}{2\pi}
\frac{\beta\bar{\omega}A_{\rho\rho}(q,\bar{\omega})} {\omega
-\bar{\omega}+i\eta}~~. \eea

\noindent From this it follows

\be A_{\rho B}(q,0)=-\int\frac{d\bar{\omega}}{2\pi} \beta
A_{\rho\rho}(q,\bar{\omega}) ~~~. \ee

\noindent Finally the element $A_{BB}(q,\omega )=0$.

We now prove the following important property of the FDRMP : If
$A_{\alpha\beta}(q,\omega )$ and $C_{\alpha\beta}(q,\omega )$ are
FDR matrix propagators then

\be D_{\alpha\beta}(q,\omega)=
\sum_{\mu\nu}A_{\alpha\mu}(q,\omega)\sigma_{\mu\nu}(q)
C_{\nu\beta}(q,\omega) \ee

\noindent is also a FDR matrix propagator. The proof is rather
direct. Look first at the response channel:

\bea D_{BB}(q,\omega) &=&
\sum_{\mu\nu}A_{B\mu}(q,\omega)\sigma_{B\nu}(q)
C_{\nu\beta}(q,\omega) \nonumber \\
&=& A_{B\rho}(q,\omega)\sigma_{\rho\rho}(q) C_{\rho B}(q,\omega)=0
~~~. \eea

\noindent Consider next the off -diagonal components

\bea D_{\rho B}(q,\omega) &=& A_{\rho B}(q,\omega)\sigma_{B\rho}(q)
C_{\rho B}(q,\omega) = A_{\rho B}(q,\omega)V(q) C_{\rho B}(q,\omega)
\nonumber \\
D_{B\rho }(q,\omega) &=& A_{B\rho }(q,\omega)V(q)
C_{B\rho}(q,\omega) \eea

\noindent It is easy to see that  $D_{\rho B}(q,\omega)=
D^{*}_{B\rho }(q,\omega).$ Next consider the diagonal component

\bea D_{\rho \rho}(q,\omega) &=& A_{\rho
\rho}(q,\omega)V(q)C_{B\rho}(q,\omega ) +A_{\rho
B}(q,\omega)V(q)C_{\rho \rho}(q,\omega ) \nonumber \\
&=& \frac{V(q)}{i\beta\omega}\left[(A_{B\rho}(q,\omega) -A_{\rho
B}(q,\omega))C_{B\rho}(q,\omega ) +A_{\rho B}(q,\omega)
(C_{B\rho}(q,\omega ) -C_{\rho B}(q,\omega ))\right] \nonumber \\
&=&\frac{V(q)}{i\beta\omega}\left[(A_{B\rho}(q,\omega)
C_{B\rho}(q,\omega ) -A_{\rho B}(q,\omega) C_{\rho B}(q,\omega
))\right]  \nonumber \\
&=&\frac{1}{i\beta\omega}\left[D_{B\rho}(q,\omega ) -D_{\rho
B}(q,\omega )\right] \eea

\noindent The above result implies that

\bea i\beta \omega D_{\rho\rho}(q,\omega ) &=& D_{B\rho}(q,\omega )
-D_{\rho B}(q,\omega ) \label{eq:616} \\
-i\beta \omega D^{*}_{\rho\rho}(q,\omega ) &=& D_{\rho B}(q,\omega )
-D_{B\rho }(q,\omega ) \label{eq:617} \eea

\noindent  Together Eqs.(\ref{eq:616}) and (\ref{eq:617}) give
$D_{\rho\rho}(q,\omega ) = D^{*}_{\rho\rho}(q,\omega )$.

Let us now consider the dressed propagators respectively denoted as
$\bar{G}$ and $\widetilde{G}$. In operator notation $\bar{G}$ and
$\widetilde{G}$ are respectively defined  as $\bar{G} = {\cal
G}\sigma{G}$ and $\widetilde{G}= {\cal G}\sigma G \sigma
{\cal G}$. Writing out explicitly the matrix forms we obtain for
$\bar{G}$ and $\widetilde{G}$ the following expressions :

\bea\bar{G}_{\alpha\beta}(q,\omega ) &=& {\cal G}_{\alpha
\mu}(q,\omega ) \sigma_{\mu\nu}(q)G_{\nu \beta}(q,\omega) \\
\widetilde{G}_{\alpha\beta}(q,\omega ) &=& {\cal G}_{\alpha
\mu}(q,\omega ) \sigma_{\mu\nu}(q){G}_{\nu \delta}(q,\omega)
\sigma_{\delta\eta}(q)
{\cal G}_{\eta\beta}(q,\omega ) ~~~, \eea

\noindent From the above theorem then it follows that both $\bar{G}$
and $\widetilde{G}$ are FDRMP. This holds since each of the $G$,
${\cal G}$, and $G^{(0)}$ satisfies the conditions of being a FDRMP.

\begin{figure}
\label{cfig1}
\includegraphics[width=16cm]{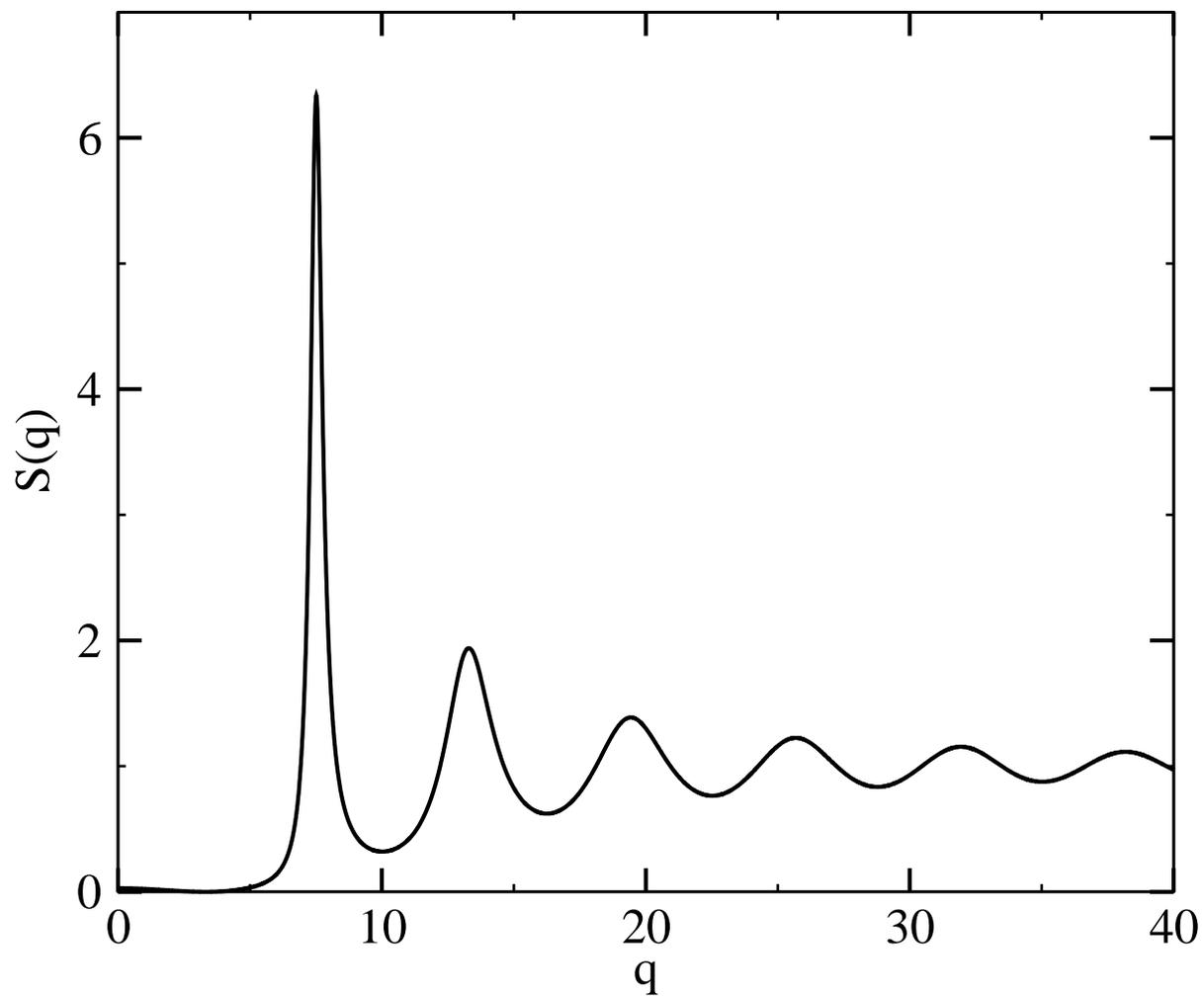}
\caption{The static structure factor of the hard sphere liquid at
packing fraction $\eta=.62$ as obtained from Percus-Yevick solution
with Verlet-Weiss corrections.}
\end{figure}

\begin{figure}
\label{cfig2}
\includegraphics[width=16cm]{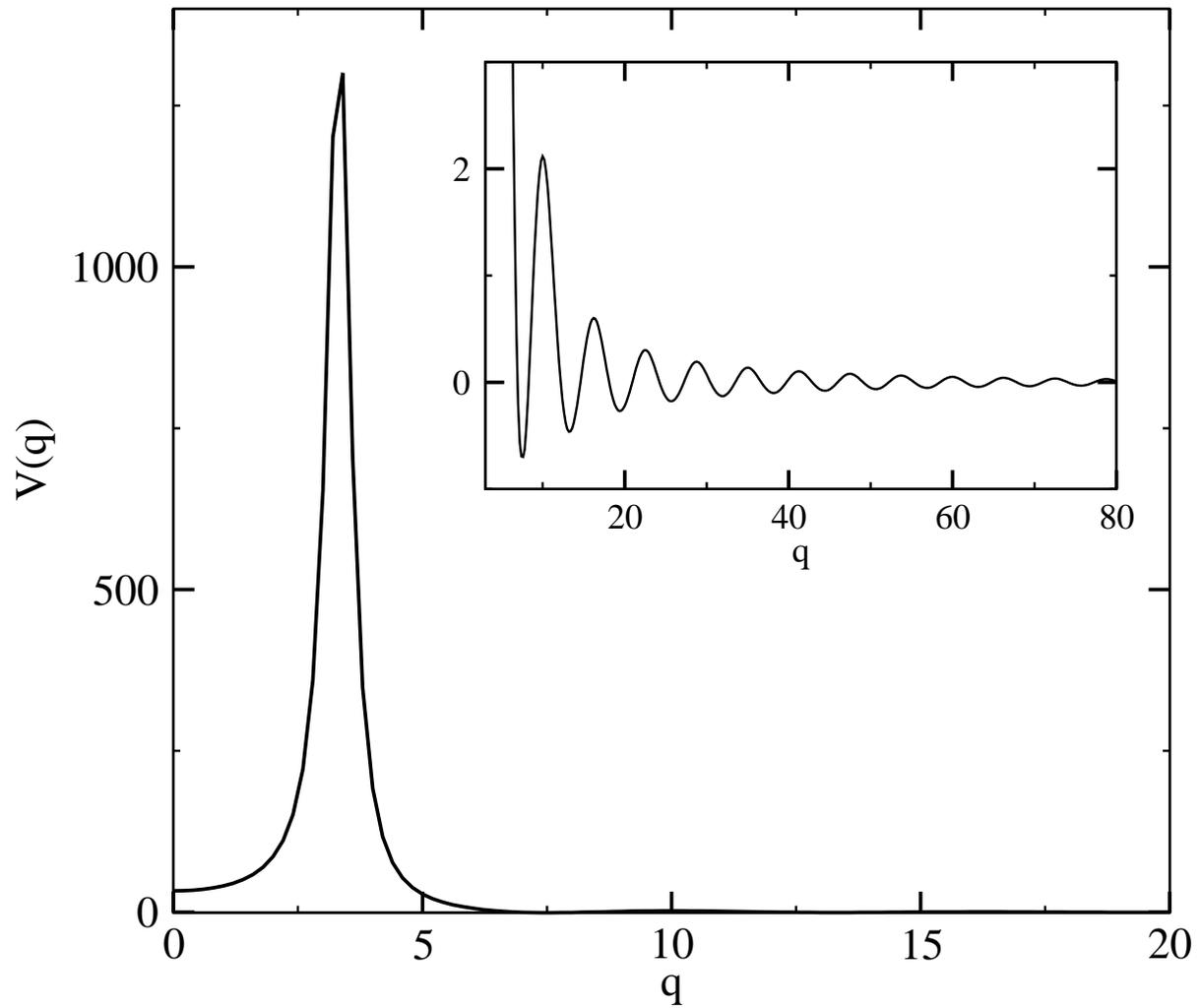}
\caption{The renormalized potential $\tilde{V}(q)$ corresponding to
the structure factor displayed in Fig. 1. The inset shows the
oscillations at large wave vectors on a enlarged scale.}
\end{figure}

\begin{figure}
\label{cfig3}
\includegraphics[width=16cm]{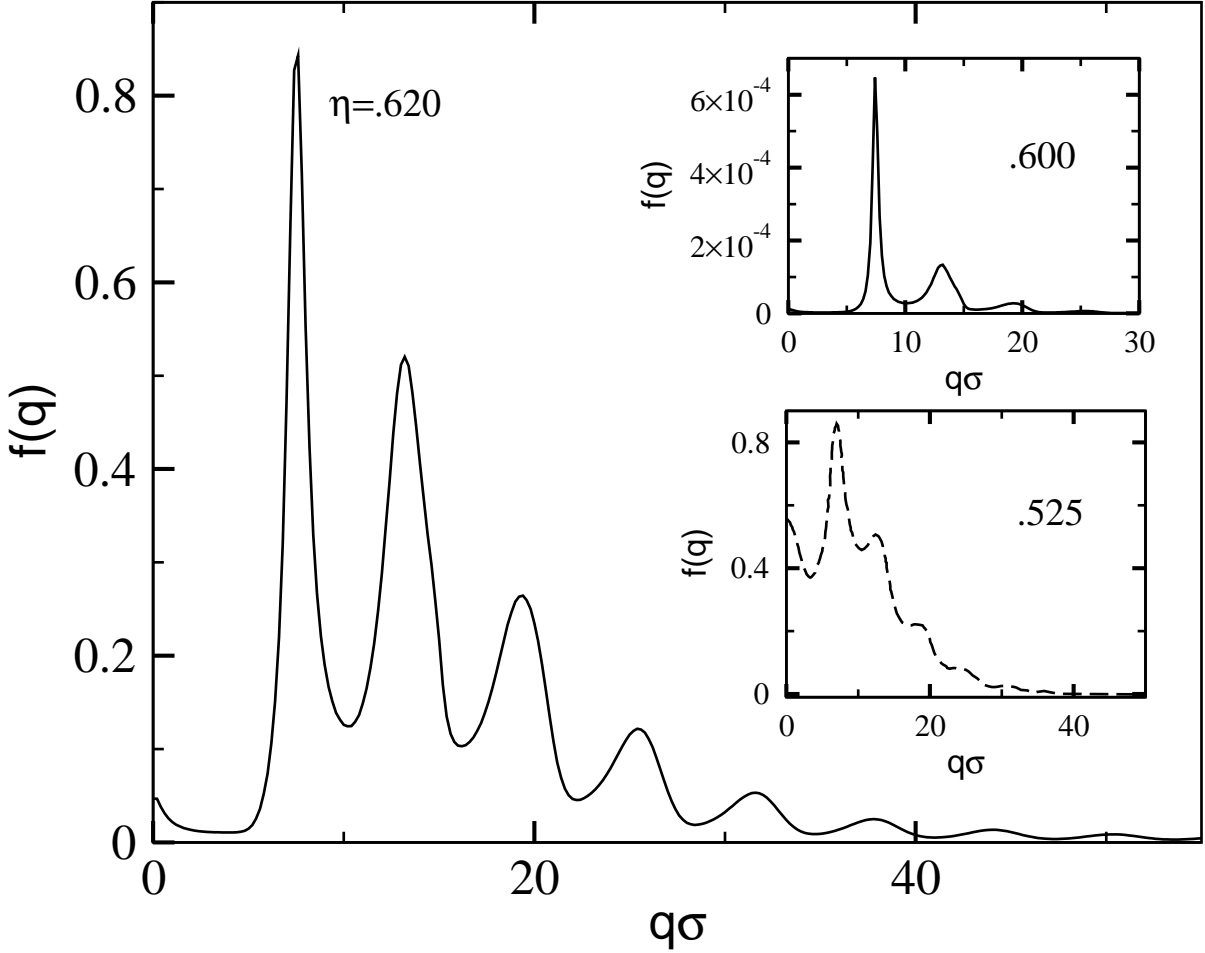}
\caption{The non ergodicity parameter $f(q)$ vs. $q\sigma$ at
packing fraction $\eta=.62$. The upper inset shows the same quantity
at $\eta=.60$ indicating ergodic behavior. The lower inset shows the
$f(q)$ vs. $q\sigma$ at the ENE transition point $\eta=.525$ in the
earlier MCT of Ref. 10}
\end{figure}


\begin{thebibliography}{99}
\bibitem{FDRSD} Similar relations have been found in the SD case.
\bibitem{MSR}
MSR conjugate variables and  actions are discussed in Appendix A in FTSPD.
\bibitem{TDDC} In different contexts the kinetic kernels have
different names.  In the field theory context the kernels are
typically called self-energies, in the kinetic theory case,
where the analysis is in terms of retarded quantities, the kernel is
called a memory function, and in the general case it can be called
a dynamic direct correlation function.
\end{thebibliography}
\end{document}